\documentclass[prr,twocolumn,showpacs,superscriptaddress,nofootinbib]{revtex4}

\usepackage{amssymb}
\usepackage{color}
\usepackage{amsmath}
\usepackage{amstext}
\usepackage{latexsym}
\usepackage{graphicx}
\usepackage[usenames,dvipsnames]{xcolor}
\usepackage[colorlinks=true,citecolor=Blue,linkcolor=RubineRed,urlcolor=Blue]{hyperref}

\def\la{\langle}
\def\ra{\rangle}

\newcommand{\beq}{\begin{equation}}
\newcommand{\eeq}{\end{equation}}
\newcommand{\beqa}{\begin{eqnarray}}
\newcommand{\eeqa}{\end{eqnarray}}

\begin{document}

\title{Distribution of Kinks in an Ising Ferromagnet After Annealing \\and the Generalized Kibble-Zurek Mechanism}

\author{Jack J. Mayo}
\affiliation{Donostia International Physics Center,  E-20018 San Sebasti\'an, Spain}
\affiliation{University of Groningen, 9712 CP Groningen, Netherlands}
\affiliation{Korteweg-de Vries Institute for Mathematics, University of Amsterdam, Science Park 105-107, 1098 XG Amsterdam, Netherlands}
\author{Zhijie  Fan}
\affiliation{Department of Physics, University of Virginia, Charlottesville, VA 22904, USA}
\author{Gia-Wei  Chern}
\affiliation{Department of Physics, University of Virginia, Charlottesville, VA 22904, USA}
\author{Adolfo del Campo}
\affiliation{Department  of  Physics  and  Materials  Science,  University  of  Luxembourg,  L-1511  Luxembourg, Luxembourg}
\affiliation{Donostia International Physics Center,  E-20018 San Sebasti\'an, Spain}
\affiliation{IKERBASQUE, Basque Foundation for Science, E-48013 Bilbao, Spain}
\affiliation{Department of Physics, University of Massachusetts, Boston, MA 02125, USA}

\begin{abstract}
We consider the annealing dynamics of a one-dimensional Ising ferromagnet induced by a temperature quench in finite time.
In the limit of slow cooling,  the asymptotic two-point correlator is analytically found under Glauber dynamics, and the distribution of the number of kinks in the final state is shown to be consistent with a Poissonian distribution. The mean kink number, the variance, and the third centered moment take the same value and obey a universal power-law scaling with the quench time in which the temperature is varied. The universal power-law scaling of cumulants is corroborated by numerical simulations based on Glauber dynamics for moderate cooling times away from the asymptotic limit,  when the kink-number distribution takes a binomial form.  We analyze the relation of these results to physics beyond the Kibble-Zurek mechanism for critical dynamics,  using the kink number distribution to assess adiabaticity and its breakdown.
We consider linear, nonlinear, and exponential cooling schedules, among which the latter provides the most efficient shortcuts to cooling in a given quench time.
The non-thermal behavior of the final state is established by considering the trace norm distance to a canonical Gibbs state.

\end{abstract}
\maketitle

\section{Introduction}
\label{sec:intro}
Nonequilibrium phenomena occupy a prominent role at the frontiers of physics, where few and highly-valuable
paradigms are able to provide a description making use of equilibrium properties.
Notable examples include linear response theory and the fluctuation-dissipation theorem \cite{Kubo66}, fluctuation theorems and work relations valid far from equilibrium \cite{Jarzynski07,Seifert12}, and the Kibble-Zurek mechanism \cite{Kibble76a,Kibble76b,Zurek96a,Zurek96c}. 
We focus on the latter, as it provides a framework to analyze the course of a phase transition and the breakdown of adiabatic dynamics leading to the formation of topological defects.
In this context, the system of interest exhibits different collective phases as a control parameter is varied across a critical value. 
This parameter is the temperature in thermal phase transitions but can be identified by other quantities such as a magnetic field, or the density of particles in the system.
The crossing of a continuous phase transition is characterized by the divergence of the (equilibrium) relaxation time in the neighborhood of the critical point, known as critical slowing down. As a result, whenever the phase transition is driven in a finite quench time $\tau_Q$, adiabaticity is broken \cite{DZ14}.

A scenario of spontaneous symmetry breaking is characterized by the presence of a manifold of degenerate ground states in the low-symmetry phase of the system. During the course of the phase transition, causally disconnected regions of the system may single out different ground states, leading to the formation of domains and the creation of topological defects at the resulting interfaces. A familiar example in this context is the cooling of a paramagnet below the Curie temperature, resulting in domains with a homogenous local magnetization and separated by domain walls.
According to the Kibble-Zurek mechanism (KZM) the mean number of defects decays as a function of the time scale in which the transition is crossed. Specifically, a universal power-law scaling is predicted when the control parameter is driven linearly in time. 
The finite-time cooling of an Ising ferromagnet has been used as a paradigmatic testbed to explore KZM physics \cite{Suzuki09,Krapivsky10,Suzuki11,Jeong20,priyanka2020slow}, which provides useful heuristics in adiabatic quantum optimization and quantum annealing \cite{Kadowaki98,Farhi01,DasBook,Das08,Albash18}.
Generalizations of KZM have been established that account for disorder \cite{Dziarmaga06,Caneva07},  nonlinear driving protocols \cite{Diptiman08,Barankov08,GomezRuiz19} as well as inhomogeneous systems \cite{KV97,ZD08,Zurek09,Dziarmaga10,CK10,ions1,DRP11,DKZ13,GomezRuiz19,Sadhukhan20,Sinha20}, see Refs. \cite{Polkovnikov11,DZ14} for a review.  These developments have inspired novel protocols in adiabatic quantum computation \cite{ZD08,Rams16,Susa18,Mohseni18,Susa18b,Adame20}.
While the early formulation of KZM was focused on classical systems, following decades of research, the applicability of the KZM in the quantum domain has been established by a combination of analytical, numerical, and experimental studies \cite{Dziarmaga10,Polkovnikov11,DZ14}.

Beyond the mean number of kinks, one may wonder whether the full kink number distribution exhibits universal behavior. The latter is directly accessible in many experiments and can be as well probed via single-qubit interferometry \cite{Xu19}.
The kink number distribution has recently been shown to exhibit signatures of universality beyond the KZM in a family of models known as quasi-free fermion systems, that include paradigmatic instances such as the one-dimensional transverse-field Ising and XY models, and the Kitaev chain  \cite{delcampo18}.  In particular, not only the mean number of defects but as well the variance, third centered-moment, and any cumulant of the kink number distribution of higher order have been shown to scale following a  universal power-law with the quench time \cite{Cincio07,delcampo18,Cui2019}.  This prediction has been experimentally explored using a trapped-ion for the quantum simulation of critical dynamics in momentum space \cite{Cui2019}. Universal features of kink number statistics in the one-dimensional transverse-field quantum Ising model have also been reported using D-Wave quantum annealers as quantum simulators \cite{Bando20}.   It is thus natural to wonder whether the distribution of topological defects in classical systems is as well universal. Indeed, a framework to account for the distribution of topological defects generated across a classical continuous phase transition has been put forward and predicts a binomial distribution, in agreement with numerical simulations for the time-dependent Ginzburg-Landau theory  \cite{GRMdC19}. In higher dimensional systems,  further evidence for the presence of universality in the full counting statistics of topological defects has been provided by the study of the vortex number distribution in a newborn holographic superconductor, which was predicted to be Poissonian \cite{delcampo2021universal}. At the time of writing, experimental evidence of Poissonian  vortex statistics has been reported after cooling of an atomic Bose gas into a superfluid in finite time \cite{goo2021defect}.

In this work, we characterize the exact kink-number distribution of a one-dimensional classical ferromagnet cooled in finite time. 
Specifically, we consider the one-dimensional Ising model with no magnetic field and evolving under Glauber dynamics.
The mean number of defects in this setting has been studied by Krapivsky \cite{Krapivsky10}, see as well \cite{Jeong20,priyanka2020slow} for related work.

Here, using a ring topology endowed with translational invariance, the kink number distribution is studied.
 We calculate explicit expressions for the general two-point correlator for spins separated by a lattice-point distance $n$ in the same limit.
The first three cumulants of the kink number distribution are explicitly shown to be equal and described by a universal power-law with the quench time, indicating that the slow cooling of an Ising ferromagnet under the Glauber dynamics yields a Poissonian kink-number distribution. The relevance of these findings to finite-annealing times is verified by numerical simulations, in which we consider three different families of cooling schedules: linear, nonlinear, and exponential quenches.

From the outset, we note that the only critical behavior of a one-dimensional Ising ferromagnet is exhibited at zero temperature.  A cooling schedule cannot possibly involve the crossing of the critical point considered in the original studies of KZM. However, the KZM prediction can be extended to account for  ``half-quenches'' ending at the critical point \cite{DZ06,Chandran12,Keim15,delcampo15,BD18}. At the same time, the finite-temperature treatment endows the dynamic with coarsening. The nonequilibrium dynamics is thus governed by a coexistence of KZM universality and coarsening. Their contribution can generally be discriminated by considering the time scales involved. In some instances,  such as the artificial spin ice \cite{Libal20}, the discrimination is not possible. 
However, for the one-dimensional Ising ferromagnet with non-conserved order-parameter (magnetization), domain growth due to coarsening scales with the square root of the time of evolution.
In our study, we can uniquely identify signatures of critical scaling for different kinds of quenches (linear, nonlinear, exponential, etc.), ruling out the effects of coarsening.

The equilibrium critical behavior of the one dimensional Ising ferromagnet is peculiar in that it does not exhibit a power-law divergence of the correlation length as a function of the proximity to the critical point, known in higher dimensional continuous phase transitions. As a result, the correlation length critical exponent $\nu$ is not well-defined. However, we shall see that critical scaling with the finite driving time governs the cumulants of the kink distribution as it does in the generalized KZM.

\section{Kinks distribution in an Ising ferromagnet}
\label{sec:another}

A one dimensional Ising ferromagnet in the absence of an external magnetic field is described by the Hamiltonian $\mathcal{H}=-\sum_{j}J_{ij}\sigma_i\sigma_j$ where the spin at a site $j$ can take any of the two values $\sigma_j=\pm1$. The ferromagnetic character stems from $J_{ij}\geq0$. 
We first discuss the distribution function of the number of kinks and its characteristic function. Given a one-dimensional spin chain, the number of kinks in a given configuration can be studied via  the number operator
\begin{equation}
\hat{\mathcal{N}}=\frac{1}{2}\sum_{i=1}^{N}\left(1-\sigma _{i}\sigma _{i+1}\right) ,
\label{K}
\end{equation}%
which can take integer value $k\in \lbrack 0,N]$. We assume periodic boundary conditions (in the case of an open chain, the upper limit of the sum is $N-1$ instead of $N$ and $k\in[0,N-1]$). We shall be interested in the distribution of the number of kinks
\begin{equation}
P(n)=\left\langle \delta (\hat{\mathcal{N}}-n)\right\rangle=\mathrm{Tr}[\varrho\delta (\hat{\mathcal{N}}-n)],
\end{equation}%
where $\varrho$ denotes the state of the system.

For its characterization, we shall resort to the characteristic function
\begin{equation}
\widetilde{P}(\theta )=\langle e^{i\theta \hat{\mathcal{N}}}\rangle.
\end{equation}%
As the kink number takes integer values, using the Fourier transform yields
\begin{equation}
P(n)=\frac{1}{2\pi }\int_{0}^{2\pi }d\theta \widetilde{P}(\theta )e^{-in\theta}.
\end{equation}%
The kink distribution is accessible in experiments and can be measured, e.g., via single-qubit interferometry \cite{Xu19}.
We shall focus on the distribution of kinks in the nonequilibrium state resulting from cooling the Ising ferromagnet in a finite time. For its analysis, it will prove useful to use the cumulants $\kappa _{j}$ ($ j\in\mathbb{N}$) of the distribution. The cumulant generating function of the kink number distribution  is the logarithm of the characteristic function and admits the expansion

\begin{equation}
\label{cumulantexp}
\log \widetilde{P}(\theta)=\sum_{j=1}^{\infty }\kappa _{j}\frac{(i\theta )^{j}}{j!}.
\end{equation}
In particular, we shall focus on the mean given by $\kappa_1=\la \hat{\mathcal{N}}\ra$, the variance $\kappa_2=\la \hat{\mathcal{N}}^2\ra-\la \hat{\mathcal{N}}\ra^2$ and the third-centered moment 
$\kappa_3=\la (\hat{\mathcal{N}}-\kappa_1)^3\ra$.

\section{Cooling by Glauber dynamics of an Ising ferromagnet: Exact solution}\label{ExCool}

To describe the finite-time cooling of the Ising ferromagnet we shall consider its evolution under Glauber dynamics \cite{Glauber63,NaimBook}.
Specifically, we consider the nonequilibrium quenching process in which the evolution of the Ising chain is described as a reversible Markov process obeying the detailed balance condition $P_{\textrm{eq}}(\vec{\sigma})w_{i}(\vec{\sigma})=P_{\textrm{eq}}(\vec{\sigma}^{(i)})w_{i}(\vec{\sigma}^{(i)})$,  where $\vec{\sigma} = (\sigma_{1},... ,\sigma_{i},... ,\sigma_{N})$ is the current system state and $\vec{\sigma}^{(i)} = (\sigma_{1},\dots,-\sigma_{i},\dots,\sigma_{N})$ denotes the same state with the $i$th spin flipped. The probabilities $P_{\textrm{eq}}$ given by the Boltzmann distribution
\begin{equation}
\label{Boltzprob}
P_{\textrm{eq}}(\vec{\sigma})=\frac{e^{-\beta \mathcal{H}(\vec{\sigma})}}{\mathcal{Z}},    
\end{equation}
where the partition function is given by $\mathcal{Z}=\sum_{\{\sigma_i=\pm1\}} e^{-\beta \mathcal{H}(\vec{\sigma})}$.
The flipping rate $w_{i}(\vec{\sigma})$ of spin $i$ is obtained by direct substitution of Eq. (\ref{Boltzprob}) into the detailed balance condition:
\begin{align}
\label{glauberw}
\frac{w_{i}(\vec{\sigma})}{w_{i}(\vec{\sigma}^{(i)})}&=\frac{P_{\textrm{eq}}(\vec{\sigma})}{P_{\textrm{eq}}(\vec{\sigma}^{(i)})} \\ &=\frac{e^{-\beta \sigma_{i}\sum_{j}J_{ij}\sigma_{j}}}{e^{\beta \sigma_{i}\sum_{j}J_{ij}\sigma_{j}}} \\ &=\frac{1-\sigma_{i}\textrm{tanh}(\beta\sum_{j \in \langle i \rangle}J_{ij}\sigma_{j})}{1+\sigma_{i}\textrm{tanh}(\beta\sum_{j\in \langle i \rangle}J_{ij}\sigma_{j})}, 
\end{align}
where the final equality is obtained by substitution of the hyperbolic identity for the exponential. 
We shall focus on the uniform coupling case with nearest neighbor interactions, i.e., $J_{ij}=J\delta_{i,i+1}$.
Equation (\ref{glauberw})  then implies the most general flipping rate in this case to be
\begin{align}
\label{glauber_acceptance_rate}
w_{i}(\vec{\sigma}) &= \frac{\alpha}{2}\left[1-\sigma_{i}\textrm{tanh}\left(2\beta J\frac{\sigma_{i-1}+\sigma_{i+1}}{2}\right)\right]\nonumber\\ &= \frac{\alpha}{2}\left(1-\gamma\sigma_{i}\frac{\sigma_{i-1}+\sigma_{i+1}}{2}\right),
\end{align}
where we have normalized the rates by setting the limit $w_{i}\rightarrow \alpha/2$ as $T\rightarrow \infty$ and defined $\gamma=\textrm{tanh}(2 \beta J)$ for convenience. 

The Glauber dynamics of Ising ferromagnets have been the subject of an extensive literature. We focus on the non-equilibrium case, in which both $\gamma = \gamma(t)$ and $\alpha=\alpha(t)$ act as control parameters of the temperature and local flipping barrier, respectively.  Making use  of translational invariance, consider  the correlator between nearest-neighbor spins $W_1= \langle \sigma_{i} \sigma_{i+1}\rangle$. Previous work by Krapivsky indicates that in the slow-cooling regime this correlator takes the form  \cite{Krapivsky10}.
\begin{equation}
W_{1} = 1 - C\tau_{Q}^{-\delta},
\end{equation}
(up to a logarithmic factor in $\tau_{Q}$) where $C$ is a constant dependent on the cooling schedule specifics, $\tau_{Q}$ is the time taken in total for the temperature to pass from an effectively infinite value to $T=0$  and the power-law exponent $\delta$ is set by the dynamic critical exponent $z$, the cooling schedule, and system dimensionality; see as well \cite{Jeong20,priyanka2020slow}. 

The study of kink statistics requires the calculation of moments $\langle\hat{\mathcal{N}}^{m}\rangle$ of the kink operator $\hat{\mathcal{N}}$, each of which involves the evaluation of expressions proportional to correlators up to and including $2m$ individual spins. In the classical Ising model, it is known that even correlators may also be decomposed into an alternating sum of two-point correlators, parameterized solely by their separation $n$ and their time dependence under translational invariance \cite{NaimBook}. In the limit of slow cooling as $W_{1}\rightarrow 1$, it is expected that the correlator $W_{n}$ of particles separated by a distance $n$ 
\begin{equation}
W_{n}=\frac{1}{N}\sum_{i=1}^{N}\sigma_{i}\sigma_{i+n}.
\end{equation}
also converges with the same power law in $\tau_{Q}$. In this limit, the dependence of correlators $W_{n}$ should linearize their functional dependence on the distance $n$. This motivates us to start considering the behavior of two-point functions of the form
\begin{equation}
\label{corrn}
W_{n}=1-nC\tau_{Q}^{-\delta}.
\end{equation}   
In the sections to come, we derive explicit asymptotic expressions for $W_{n}$ in the case of linear, algebraic, and exponential quenches from the generating function. 
We shall use these results to establish the universal form of the kink number distribution and the scaling of its cumulants with the quench time.

We also perform extensive dynamical simulations of the ferromagnetic Ising-Glauber model with various cooling schedules. The Glauber dynamics, which is equivalent to the so-called heat-bath method in Monte Carlo simulations, can be easily implemented numerically. To take into account the stochastic and local nature of the spin dynamics, at each fundamental step, a spin $\sigma_i$ that is randomly chosen from the system is to be updated according to the Glauber transition dynamics. Specifically, a random number $r$ uniformly distributed in the interval $[0, 1]$ is generated from a pseudo-random number generator. The chosen spin is flipped, i.e. $\sigma_i \to -\sigma_i$ if this random number satisfies $r < w_i(\vec\sigma)$, where $w_i(\vec\sigma)$ is given by the Glauber acceptance rate ~Eq.~(\ref{glauber_acceptance_rate}) with $\alpha$ set to~1.
To properly compare simulation results from different system sizes $N$, we define a time-step in our simulations as consisting of $N$ single spin-updates described above. The system is initialized in a random spin configuration and then cooled down by tuning the control parameter $\gamma$ at each time step according to the cooling schedule. For each cooling speed $\tau_Q$, a large number of independent cooling simulations are performed and observables are computed from instantaneous snapshots of the spin configurations.

Periodic boundary conditions are used in all our numerical simulations presented below. 
When a final configuration is generated, the kink number $\hat{\mathcal{N}}$ is measured in different configurations, obtaining the  the moments $\langle \hat{\mathcal{N}}\rangle$, $\langle \hat{\mathcal{N}}^2\rangle$, $\langle \hat{\mathcal{N}}^3\rangle$ from the Monte Carlo average. 
The cumulants are then calculated using the identities
\begin{eqnarray}
	\label{cumulant_formulas}
	& & \kappa_1 = \langle \hat{\mathcal{N}} \rangle, \nonumber\\ 
	& & \kappa_2 = \langle \hat{\mathcal{N}}^{2} \rangle - \langle \hat{\mathcal{N}} \rangle^{2}, \\ 
	& & \kappa_3=\langle\hat{\mathcal{N}}^{3}\rangle-3 \langle\hat{\mathcal{N}}\rangle \langle\hat{\mathcal{N}}^{2}\rangle + 2\langle\hat{\mathcal{N}}\rangle^{3}, \nonumber
\end{eqnarray}
 where $\langle \cdots \rangle$ denotes average over independent annealing simulations.

\subsection{Cumulant Generating Function}

Consider the probability distribution for a given spin configuration
$P(\vec{\sigma},t)$. Under Glauber dynamics, this distribution evolves according to the master equation
\begin{equation}
\label{masterequation}
\frac{\partial}{\partial t}P(\vec{\sigma},t)=
-\sum_iw_{i}(\vec{\sigma})P(\vec{\sigma},t)+\sum_iw_{i}(\vec{\sigma}^{(i)})P(\vec{\sigma}^{(i)},t).
\end{equation}
Correlation functions between different spins can be found in terms of the generating function introduced by Aliev \cite{Aliev98,Aliev00,Aliev09}
\begin{eqnarray}
\Psi(\{\eta_i\},t)&=&\left\langle\prod_i(1+\eta_i\sigma_{i})\right\rangle_{\vec{\sigma}}\nonumber\\
&=&\sum_{\vec{\sigma}}P(\vec{\sigma},t)\prod_i(1+\eta_i\sigma_{i}),
\end{eqnarray}
where $\langle \cdot \rangle_{\vec{\sigma}}$ denotes an expectation over spin realizations $\vec{\sigma}$, and $\{\eta_i\}$ is a set of Grassmann variables satisfying
\begin{equation}
\eta_i^2=0,\quad \eta_i\eta_j+\eta_j\eta_i=0.
\end{equation}
and we assume an infinite chain for simplicity.
Explicit correlation  functions can be derived from the generating function via the identity, applicable for an even number $n$ of indices $i_{j}$: 
\begin{equation}
\langle \sigma_{i_{1}} \sigma_{i_{2}} \cdots \sigma_{i_{n}}\rangle = \frac{\partial^{n}\Psi(\{ \eta_{i} \};t)}{\partial \eta_{i_{n}} \cdots \partial \eta_{i_{2}} \partial \eta_{i_{1}}}\bigg|_{\{ \eta_{i}\}=0}.
\end{equation}
After the propagation of the generating function by way of (\ref{masterequation}) in the manner described by Aliev \citep{Aliev09}, the form of $\Psi(\{ \eta_{i} \};t)$ induced by the Grassmann variables $\{\eta_i\}$ allows the explicit expression by differentiation of correlators in the form of an alternating sum of products of two-point functions. More concisely, this may be encoded as
\begin{equation}
\label{spincorr}
\frac{\partial^{n}\Psi(\{ \eta_{i} \};t)}{\partial \eta_{i_{n}} \cdots \partial \eta_{i_{2}} \partial \eta_{i_{1}}}\bigg|_{\{ \eta_{i}\}=0} = \mathrm{Pf}(W_{i_{1},i_{2},...,i_{n}}),
\end{equation} 
where $W_{i_{1},i_{2},...,i_{n}}$ is an antisymmetric $2n\times 2n$ matrix whose elements are defined in terms of the two-point correlators $W_{i_{k}i_{l}}=\langle \sigma_{i_{k}}\sigma_{i_{l}} \rangle$:
\begin{equation}
(W_{i_{1},i_{2},...,i_{n}})_{kl} =
\left\{
	\begin{array}{ll}
		W_{i_{k}i_{l}}  & \mbox{if } k<l \\
		0  & \mbox{if } k=l \\
		-W_{i_{k}i_{l}} & \mbox{if } k>l 
	\end{array}
\right. .
\end{equation}
In Eq. (\ref{spincorr}), we use the Pfaffian $\mathrm{Pf(A)}$ of a matrix with elements $a_{kl}$, defined by the alternating sum of permutations $\pi$ over the ordered list of integers $\{1,2,...,n\}$
\begin{eqnarray}
\mathrm{Pf(A)} &=& \mathrm{det}(A)^{\frac{1}{2}} \\&=& \frac{1}{2^{\frac{n}{2}}(\frac{n}{2})!}\sum_{\pi}\mathrm{sgn}(\pi)a_{\pi(1)\pi(2)}\cdots a_{\pi(n-1)\pi(n)}.\nonumber
\end{eqnarray}
The Pfaffian equivalence induces an equivalent structure to the Wick contraction for fermionic field operators. A power-series expansion of the kink number characteristic function involves these correlators
\begin{align}
\begin{split}
\widetilde{P}(\theta )=e^{\frac{i\theta N}{2}}&\left[1 +\frac{\theta}{2i}\sum_{n}\langle\sigma_{n}\sigma_{n+1}\rangle \right. \\ 
&\left. + \frac{1}{2!}\left(\frac{\theta}{2i}\right)^{2}\sum_{n,m}\langle\sigma_{n}\sigma_{n+1}\sigma_{m}\sigma_{m+1}\rangle 
+\mathcal{O}(\theta^3)
\right].
\end{split}
\end{align} 
Thus, we can formally write the characteristic function in terms of the generating function $\Psi=\Psi(\{ \eta_{i} \};t)$
\begin{equation}
\label{gfunc}
\widetilde{P}(\theta )=e^{\frac{i\theta N}{2}}\left[1 +\left (e^{\frac{-i\theta}{2}\sum_{n}\frac{\partial^{2}}{\partial \eta_{n+1}\partial \eta_{n}}}-1\right)\Psi\right]\bigg|_{\{ \eta_{i}\}=0}.
\end{equation} 
Its logarithm, $\mathrm{ln}\tilde{P}(\theta) $, is the cumulant generating function. 
Specifically, the $j$-th cumulant $\kappa_j$ of the kink number distribution can be found as
\begin{equation}
\kappa_j=
\frac{1}{i^j}\frac{d^j}{d\theta^j}\mathrm{ln}\tilde{P}(\theta)\bigg|_{\theta=0}, \quad j\in\mathbb{N}.
\end{equation}
Let us consider three first terms with $j=1,2,3$. One readily finds the mean number of kinks as the first cumulant, i.e, 
\begin{eqnarray}
\label{kappa1}
\kappa_{1} &=&\frac{1}{2}(N-\sum_{n}(\partial^{2}_{\eta_{n+1},\eta_{n}}\Psi)|_{\{ \eta_{i}\}=0})\nonumber\\
&=&\frac{1}{2}\sum_{n}(1-\langle \sigma_{n} \sigma_{n+1}^{z}\rangle)\nonumber\\
 &=&  \langle \hat{\mathcal{N}}\rangle =\frac{N}{2}(1-W_{1}).
\end{eqnarray}
Similarly, the correlator between two spins that are $n$ sites apart in the presence of translational invariance will be denoted by $W_n=\frac{1}{N}\sum_i\langle \sigma_{i} \sigma_{i+n}\rangle$.

The explicit computation of higher-order cumulants is somewhat laborious. Here, we simply quote the result for the second and third cumulant derived in the appendix \ref{appk23}.
The second cumulant equals the variance of the number of kinks and reads
\begin{eqnarray}
\label{k2w}
\kappa_2 &=& \langle \hat{\mathcal{N}}^{2} \rangle - \langle \hat{\mathcal{N}} \rangle ^{2} \\
& =& \frac{1}{4}N\left[1 - W_{1}^{2} + 2\sum_{n = 1}^{N/2}(W_{n}^{2}-W_{n+1}W_{n-1})\right].\nonumber
\end{eqnarray}
The third cumulant equals the third centered moment and its explicit computation yields
\beqa
\label{k3w}
\kappa_{3} &=&\langle (\hat{\mathcal{N}} - \langle \hat{\mathcal{N}}\rangle)^3\rangle  \\
&=& \frac{1}{4}NW_{1}\left[1-W_{1}^{2} \right. \nonumber \\  &+& \left. \sum_{n=1}^{\frac{N-1}{2}}(N+2-4n)(W_{n}^{2}-W_{n+1}W_{n-1}).\right]. \nonumber
\eeqa

At this stage, we can analyze the general features of the kink distribution in the binomial model. 
 The latter is associated with $N$ Bernoulli trials describing the presence of a kink at the interface between different spins with a success probability $p$.
 The first three cumulants of the binomial distribution are given by  $Np$, $Np(1-p)$ and $Np(1-p)(1-2p)$.
 From the expression of the mean, one can identify the probability  $p$ in terms of the spin-spin correlator as $p=\frac{1}{2}(1-W_{1})$. An analogous identification holds for $\kappa_{2}$, with (\ref{k2w}) having first term $\kappa_{2}=\frac{1}{4}(1-W_{1}^{2})$. Regarding the $\kappa_3$ the first two terms  in  (\ref{k3w}) are consistent with the binomial expression
 as $\kappa_{3}=p(1-p)(1-2p)=\frac{1}{4}W_{1}(1-W_{1}^{2})$.
We shall revisit the connection with the binomial distribution in a different framework, that of the generalized KZM, in Section \ref{Sec:GenKZM}.

To summarize this section, we have obtained exact expressions - Eqs. (\ref{kappa1}), (\ref{k2w}) and (\ref{k3w}) - for the first three cumulants of the kink number distribution in terms of the $n$-site correlator $W_n$. A crucial observation is that these equations, together with the ansatz for the leading power-law behavior of $W_n$ in Eq. (\ref{corrn}), yield, to leading order in $1/\tau_Q$
\beqa
\label{kjPoss}
\kappa_{1}=\kappa_{2}=\kappa_{3}=\frac{N}{2}C\tau_{Q}^{-\delta},
\eeqa
suggesting that the kink-statistics becomes Poissonian in this limit.
We next turn our attention to the explicit analysis for specific cooling protocols.

\section{Finite Time Cooling}
We consider an infinite ring, with an uncorrelated initial state corresponding to the high-symmetry phase satisfying $\langle \sigma_{i} \rangle_{0} = \langle \sigma_{i} \sigma_{j} \rangle_{0} = \dots =0$ and no local flipping barrier (thus $\alpha(t)=\alpha_{0}=1$). In this case, the generating function reads \cite{Aliev98,Aliev09}
\begin{equation}
\Psi(\{\eta\};t)= \mathrm{exp}\left(\sum_{-\infty<f_{1}<f_{2}<\infty}\eta_{f_{1}}\eta_{f_{2}}W_{f_{1}f_{2}}(t)\right),
\end{equation}
in terms of \cite{Aliev09}
\begin{equation}
W_{m_{1},m_{2}}=\int_{0}^{t}\mathrm{d} \tau \gamma(\tau) \mathrm{exp}\left[2(\tau-t)\right]H_{m_{2}-m_{1},1}\left(2h(t,\tau)\right),
\end{equation}
where $h(t_{2},t_{1})=\int_{t_{1}}^{t_{2}}\mathrm{d}\tau \gamma(\tau)$,  $H_{m,j}(x)=I_{m-j}(x)-I_{m+j}(x)$  and  $I_{\nu}(x)$ denotes the $\nu$th modified Bessel function of the first kind. The indices $f_{i}$ denote the (ordered) index of each spin. In the case of an infinite chain, the correlators depend only on the distance $n$ between successive spins. In addition, the successive modified Bessel functions may reduce the expression via the identity
\begin{equation}
\frac{2 \nu I_{\nu}(x)}{x}=I_{\nu-1}(x)-I_{\nu+1}(x).
\end{equation}  
Thus, evaluating at the instant where $T=0$, we may write
\begin{equation}
\label{intWn}
W_{n}=n\int_{0}^{\tau_{Q}}\mathrm{d}t \gamma(t) \mathrm{exp}\left[2\left(t-\tau_{Q}\right)\right]\frac{I_{n}\left(2h\left(\tau_{Q},t\right)\right)}{h\left(\tau_{Q},t\right)},
\end{equation}
which provides the exact integral representation of the $n$-site correlator. 

Using this result, together with those derived in the preceding section, we next describe the kink statistics resulting from different cooling schedules.
Specifically, under Glauber dynamics the flipping rate is dictated by the parameter $\gamma=\textrm{tanh}(2 \beta J)$ and we shall consider different functional forms for the variation of this parameter in time 
\cite{Krapivsky10,Suzuki11,Jeong20,priyanka2020slow}.

\subsection{Linear Quench}

\begin{figure*}
	\includegraphics[width=0.32\linewidth]{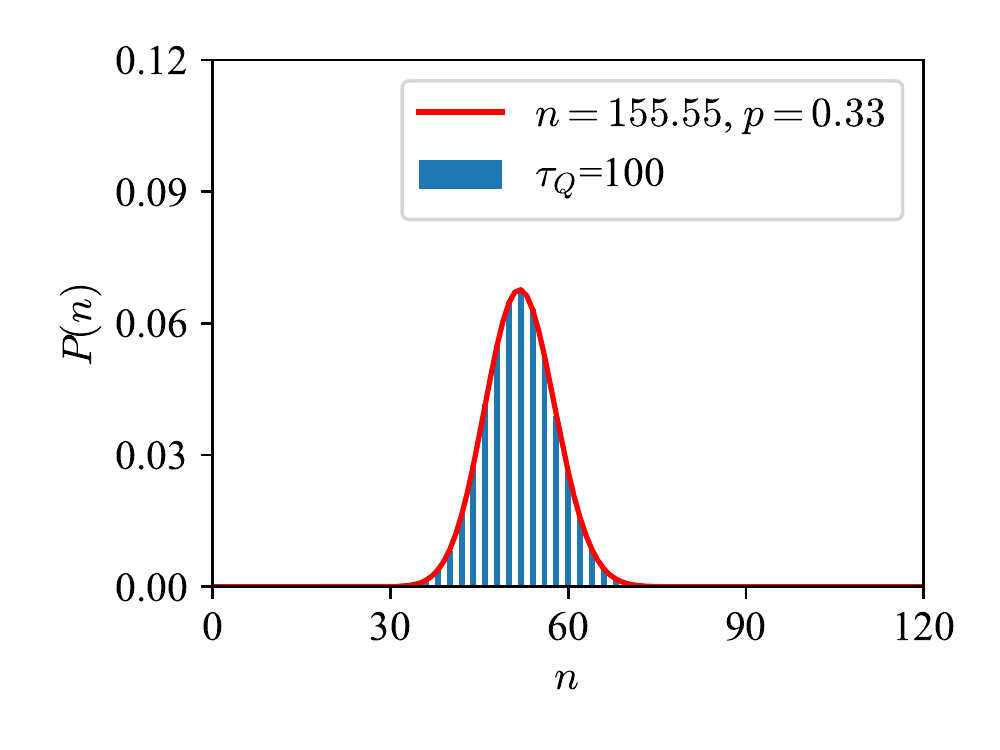}
	\includegraphics[width=0.32\linewidth]{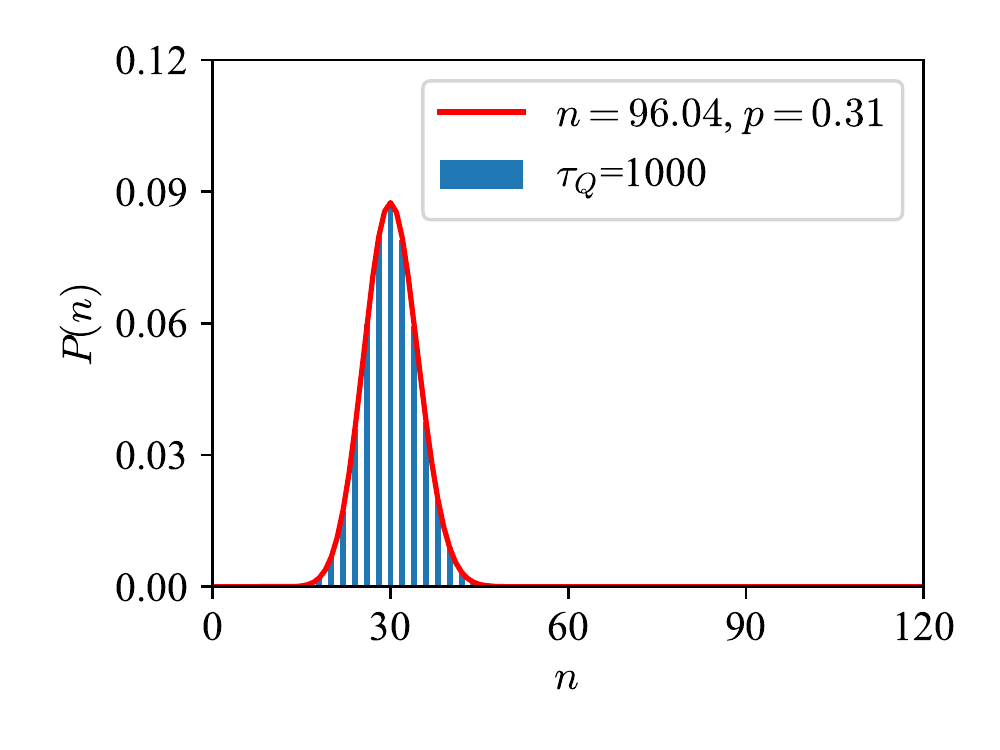}
	\includegraphics[width=0.32\linewidth]{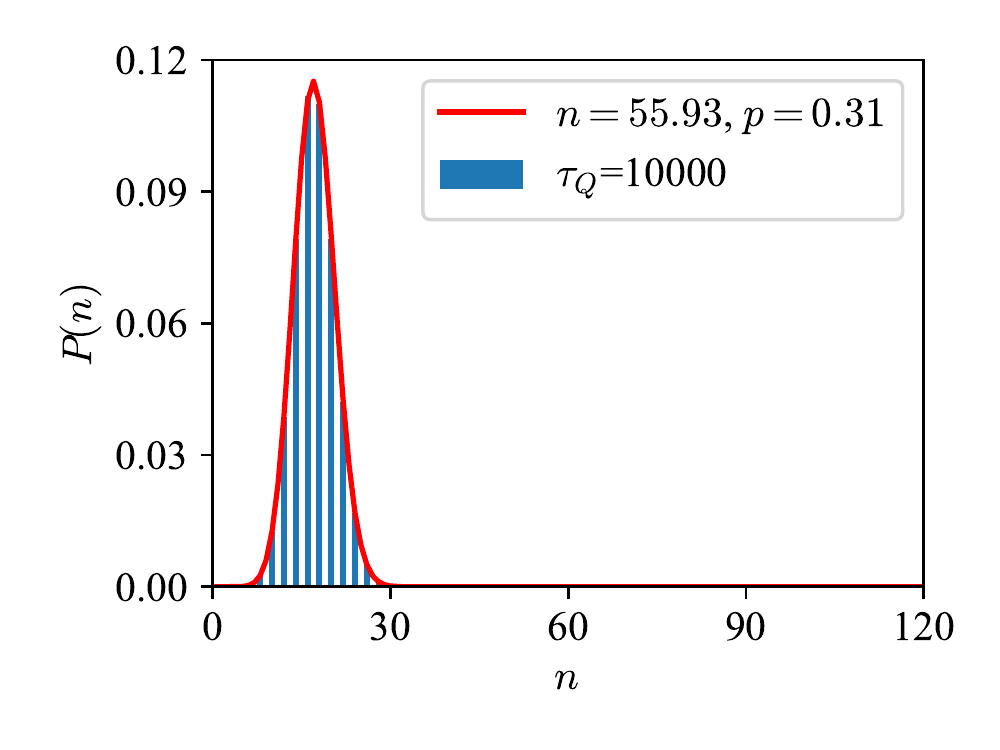}
	\caption{Kink number distribution and corresponding binomial approximation in the final nonequilibrium state of an Ising ferromagnet that is driven at different rates by a linear cooling schedule. 
	The system size is $N=500$ and the kink number distributions are obtain from $M=500000$ independent Glauber dynamics simulations.}
		\label{fig:kink-dist-linear}
\end{figure*}

\begin{figure}
	\includegraphics[width=\linewidth]{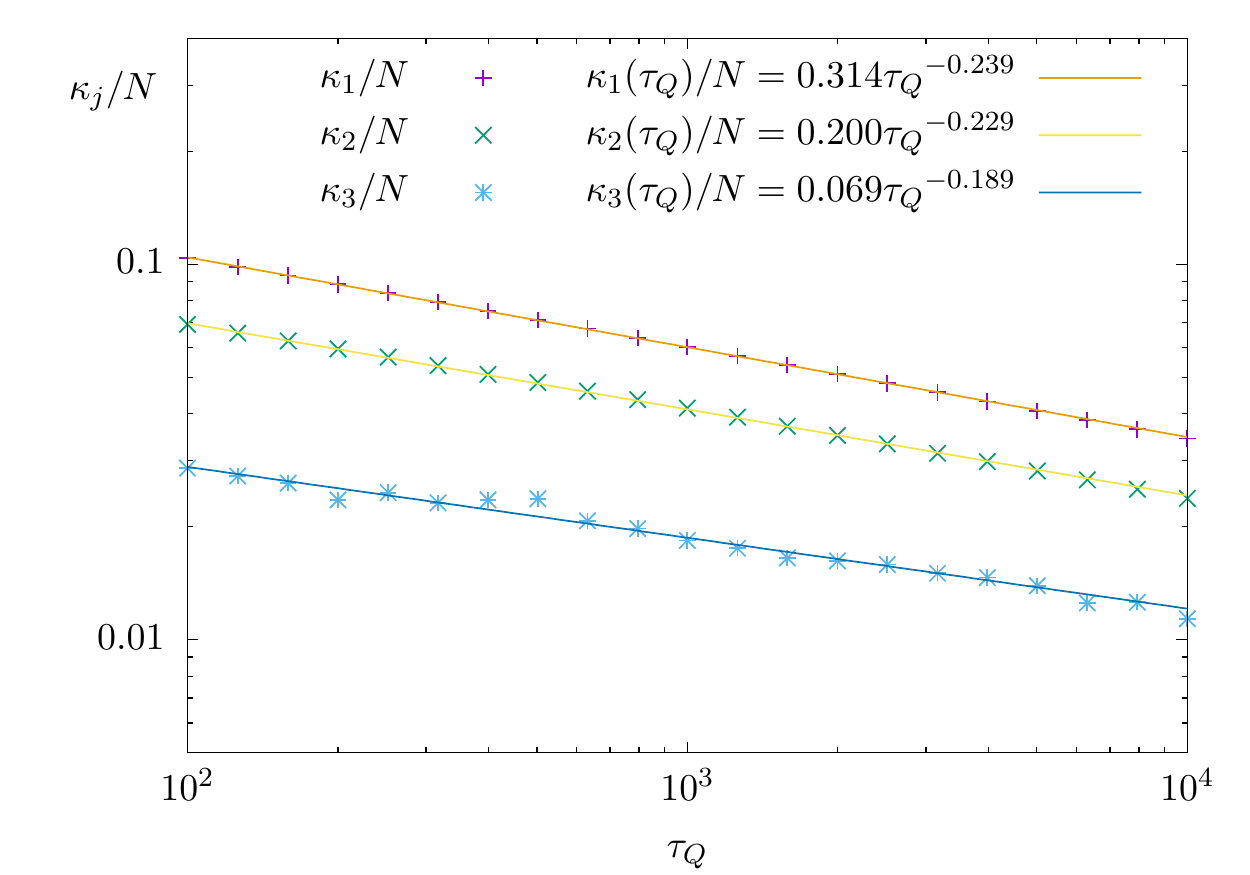}
	\includegraphics[width=\linewidth]{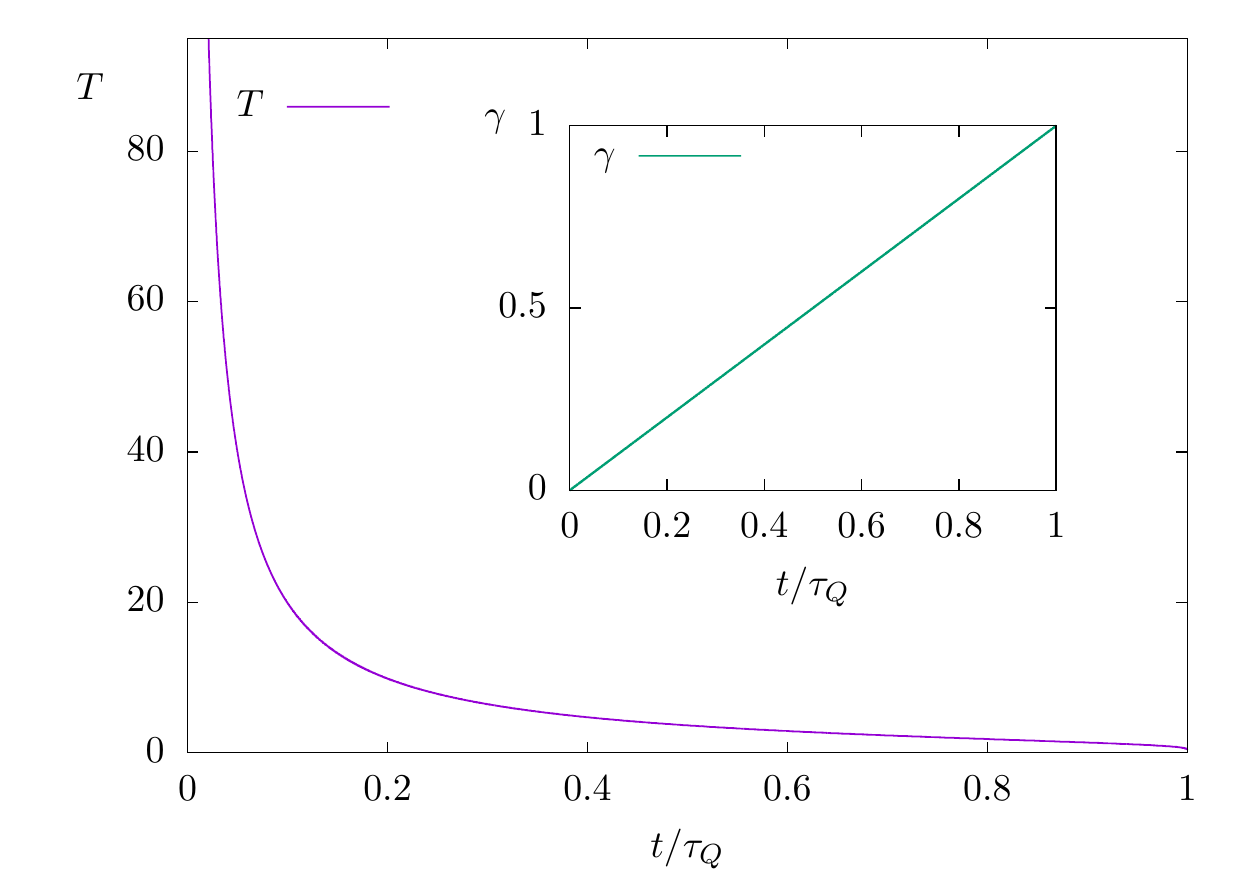}
	\caption{First three cumulants scaling of linear cooling schedules (left panel) and the cooling schedule (right panel). The system size is $N= 500$. For all three cumulants, each data point is obtained by averaging over $M = 500000$ independent Glauber dynamics simulations.}
		\label{fig:cumulants_linear}
\end{figure}

For a linear cooling schedule, we consider 
 \begin{equation}
 \label{linear_cooling_schedule}
 \gamma(t) = t/\tau_{Q},
 \end{equation}
where $\tau_{Q}$ denotes the total time taken to cross from the initial condition to the $T=0$ state. 
After evaluation at $t=\tau_{Q}$, when the critical point is reached, the integral (\ref{intWn}) reduces to
\begin{equation}
\label{linearintegral}
W_{n}(\tau_{Q})=n\int_{0}^{\tau_{Q}}\mathrm{d}\eta e^{-\eta-\frac{\eta^{2}}{4\tau_{Q}}}\frac{I_{n}(\eta)}{\eta},
\end{equation}
where we have defined $\eta = (\tau_{Q}^{2}-t^{2})/\tau_{Q}$ for convenience, and used the approximation to the exponential factor
\begin{equation}
e^{2\tau_{Q}\left(\sqrt{1-\frac{\eta}{\tau_{Q}}}-1\right)}\rightarrow e^{-\eta-\frac{\eta^{2}}{4\tau_{Q}}},
\end{equation}
justified by the exponential rolloff of the contribution for $\eta = O(\sqrt{\tau_{Q}}) $. Asymptotic solution of the integral (\ref{linearintegral}) can be exactly computed as shown in Appendix \ref{2PCorrApp} and yields
\begin{equation}
\label{corrlin}
W_{n} = 
1-\frac{n}{\sqrt{\pi}\tau_{Q}^{\frac{1}{4}}}\left[ \Gamma\left(\frac{3}{4}\right)-\frac{(4n^2-1)}{48\sqrt{\tau_{Q}}}\Gamma\left(\frac{1}{4}\right)+\mathcal{O}\left(\frac{1}{\tau_Q}\right)\right].
\end{equation}
The above result in conjunction with our results for $\kappa_{1}$ in Eq. (\ref{kappa1}), yields
\begin{equation}
\label{kappa1lin}
\kappa_{1} = \frac{N\Gamma\left(\frac{3}{4}\right)}{2\sqrt{\pi}\tau_{Q}^{\frac{1}{4}}}-
\frac{N\Gamma\left(\frac{1}{4}\right)}{32\sqrt{\pi}\tau_{Q}^{\frac{3}{4}}}+\dots
\end{equation}
By comparing the amplitude of the leading and subleading terms, one concludes that the power-law behavior sets in for quench times
\begin{equation}
\label{tau1lin}
\tau_{Q}^{(1)} \gg \frac{1}{256}\frac{\Gamma\left(\frac{1}{4}\right)^{2}}{\Gamma\left(\frac{3}{4}\right)^{2}}=0.03419\dots
\end{equation}
where the superindex indicates that this time scale characterizes the first cumulant.

In agreement with Eq. (\ref{kjPoss}), to leading order in a $1/\tau_Q$ expansion, we further find
\begin{equation}
\label{kappaqlin}
\kappa_{1}=\kappa_{2}=\kappa_{3}=\frac{N\Gamma(\frac{3}{4})}{2\sqrt{\pi}\tau_{Q}^{\frac{1}{4}}},
\end{equation}
which suggests that the distribution becomes Poissonian distribution in the limit of arbitrarily slow cooling.

Fig.~\ref{fig:kink-dist-linear} shows the probability distribution functions of kink number $P(n)$ obtained from the Glauber dynamics simulations for finite quench times. 
It is worth noting that the distribution of kinks is well described by binomial distribution $B(n, p)$ with parameters $n=\kappa_1p$, $p=1-\kappa_2/\kappa_1$. 
Fig.~\ref{fig:cumulants_linear} shows the scaling of the first three cumulants as a function of the annealing time. 
Numerically, the cumulants are calculated by averaging the recorded kink numbers using Eqs.~(\ref{cumulant_formulas}).   
Our numerical results clearly show that the first two cumulants follow a power-law dependence with the annealing rate $\tau_Q$. A relatively larger fluctuation can be seen in the data points of the third cumulant $\kappa_3$, which is expected due to the enhanced statistical error in the numerical calculation of higher-order moments. Nonetheless, the trend of $\kappa_3$ still roughly follows the power law. The power-law exponents obtained from nonlinear least-squares fitting are $0.239\pm0.001$, $0.229\pm0.001$, and $0.189\pm0.007$ for the first three cumulants. These values are close to the theoretically predicted value $1/4$  in Eq.~(\ref{kappaqlin}) but exhibit some deviations from it. 
Thus, only the first cumulant is governed by the leading  $1/\tau_Q^{1/4}$ term in this range, while the subleading corrections are important for $\kappa_2$ and $\kappa_3$.

\subsection{Nonlinear Algebraic Quench}

\begin{figure*}[t]
	\includegraphics[width=0.32\linewidth]{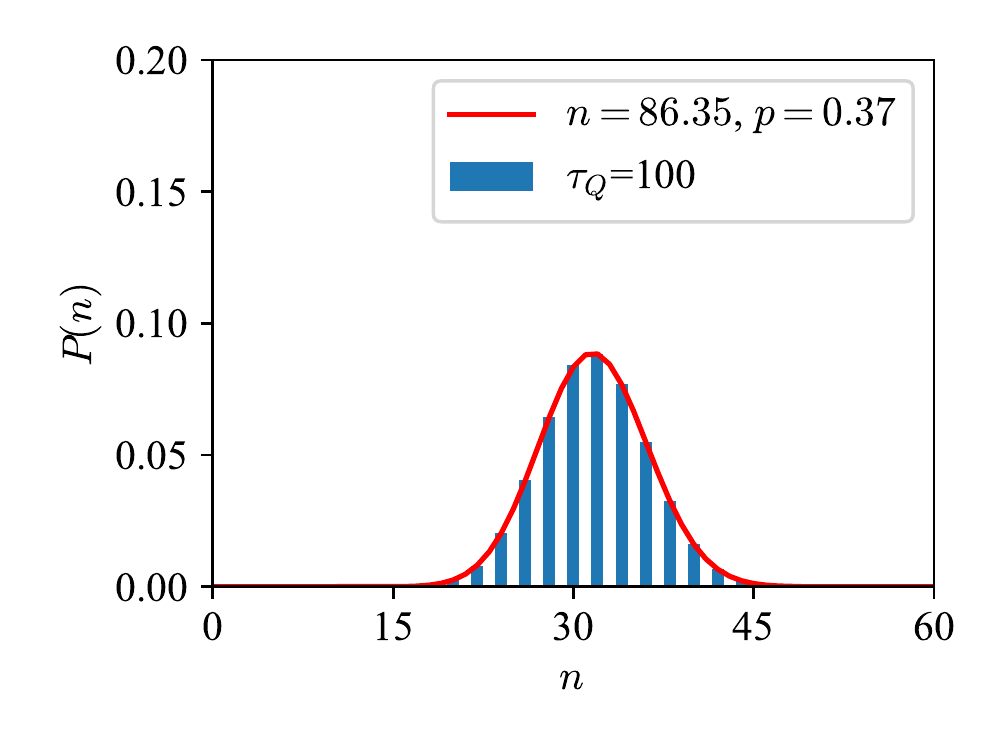}
	\includegraphics[width=0.32\linewidth]{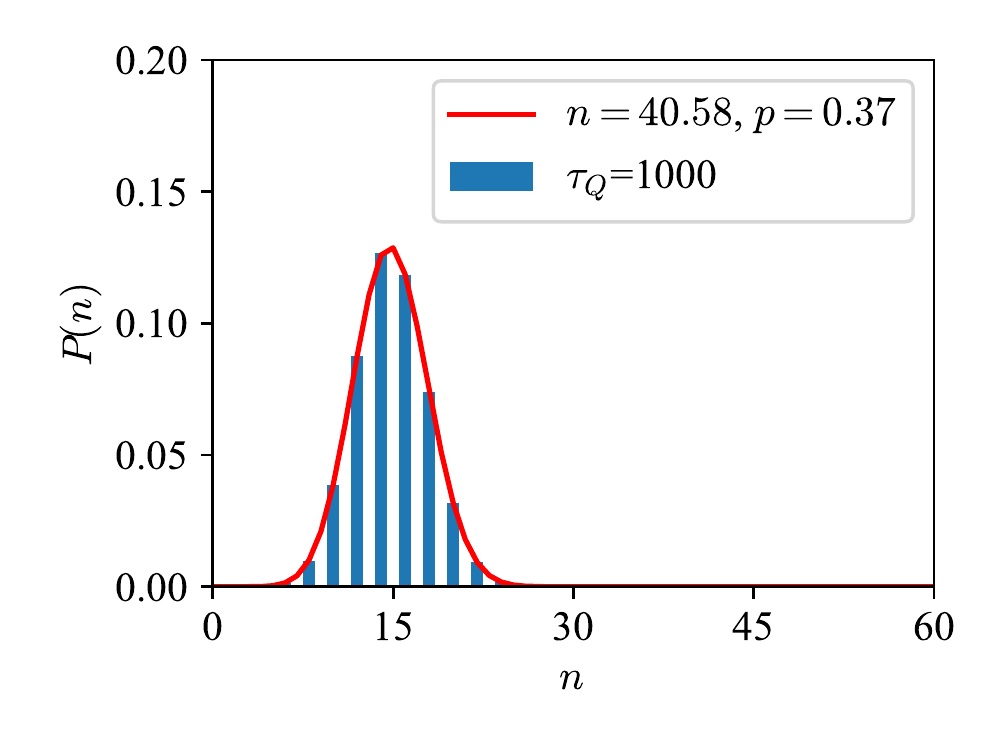}
	\includegraphics[width=0.32\linewidth]{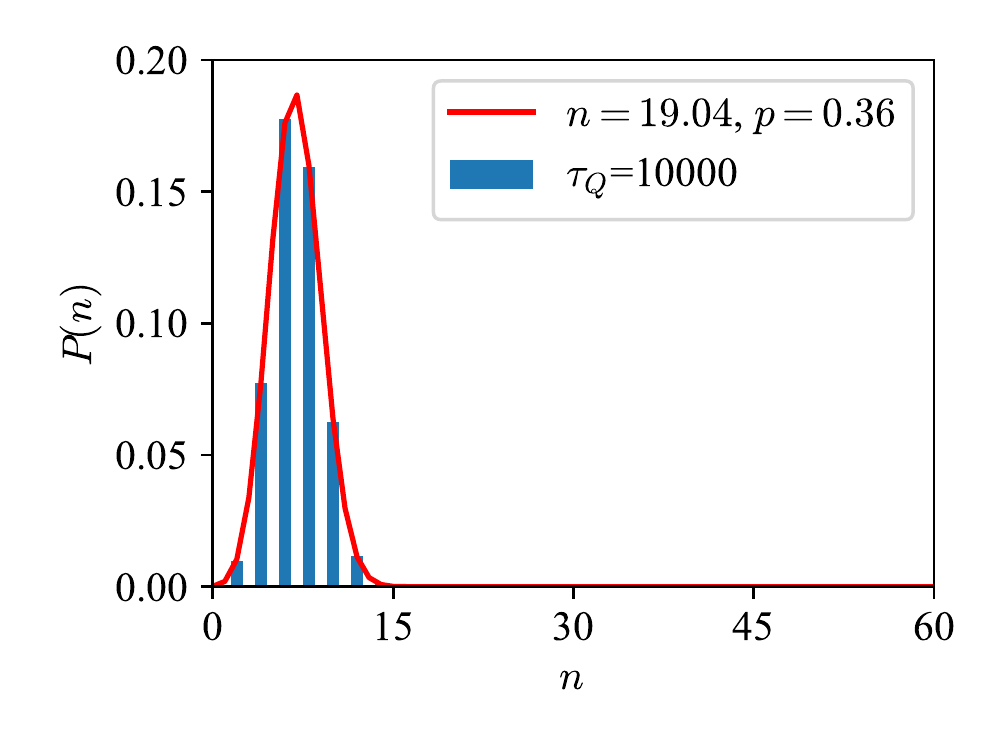}
	\caption{Kink number distribution and corresponding binomial distributions at different cooling rate for algebraic cooling schedules. The system size is $N=500$ and the kink number distributions are obtain from $M=500000$ independent Glauber dynamics simulations.}
		\label{fig:kink-dist-nonlinear}
\end{figure*}

\begin{figure}[h]
	\includegraphics[width=\linewidth]{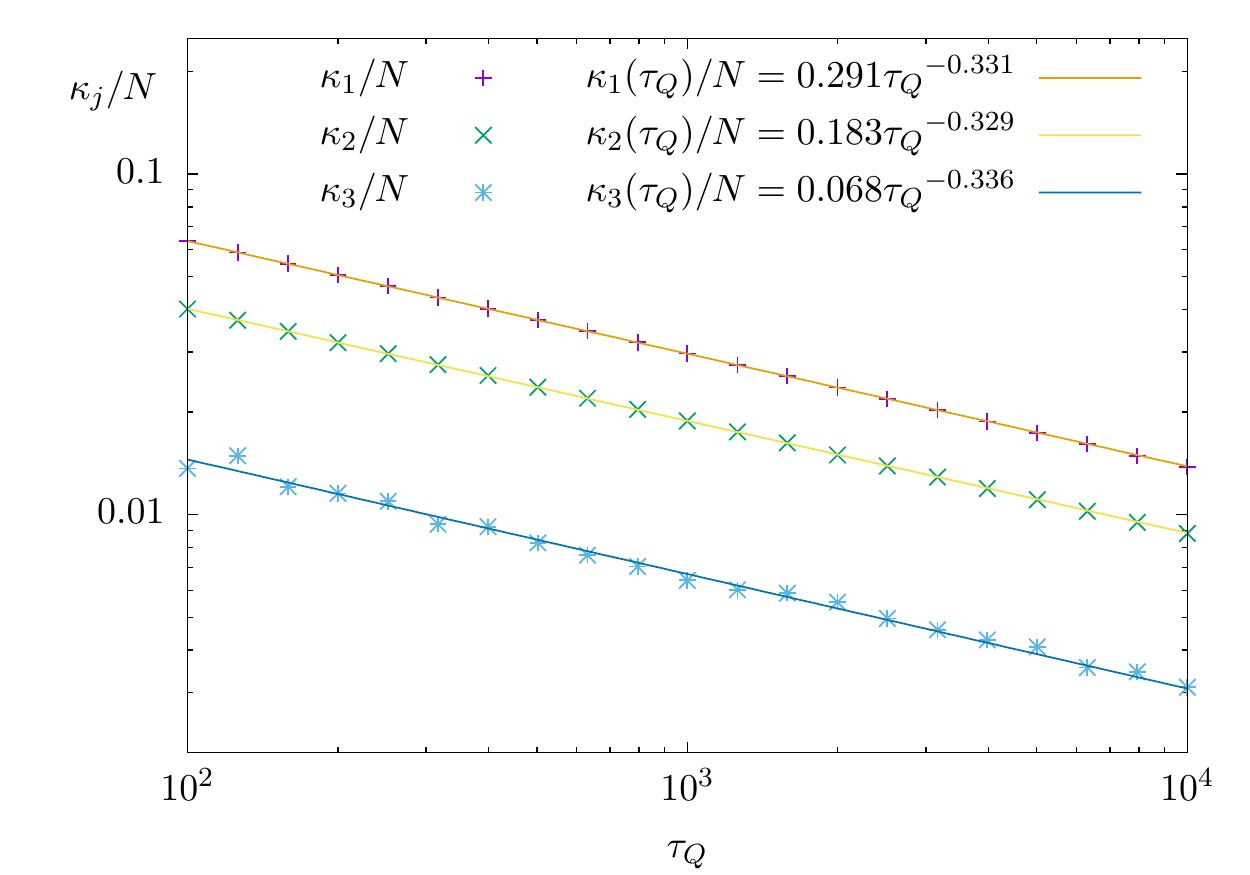}
	\includegraphics[width=\linewidth]{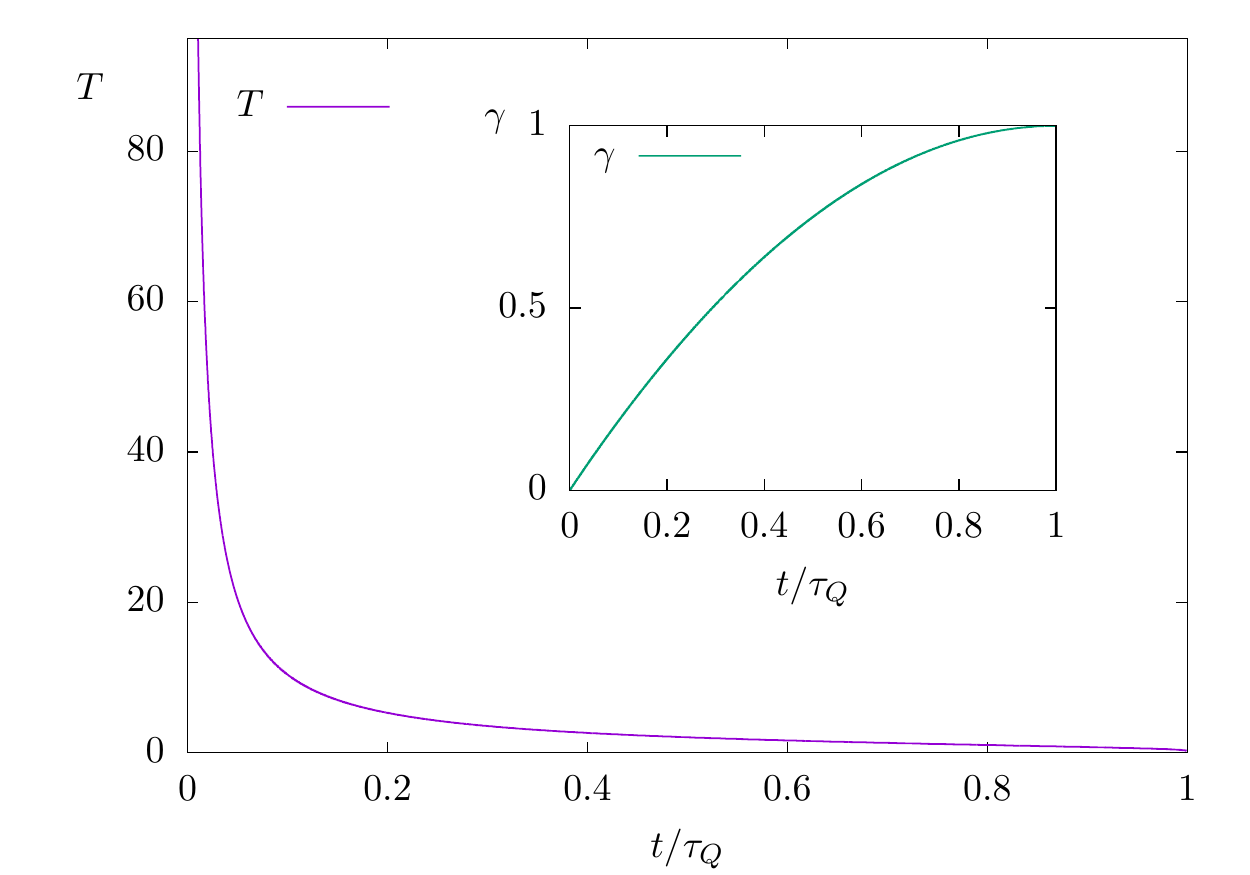}
	\caption{First three cumulants scaling of algebraic cooling schedules (left panel) and the cooling schedule (right panel). The system size is $N= 500$. For all three cumulants, each data point is obtained by averaging over $M = 500000$ independent Glauber dynamics simulations.}
		\label{fig:cumulants_nonlinear}
\end{figure}

The nonlinear passage across a critical point has been proposed to suppress the mean number of defects generated in a phase transition, as it yields a power-law dependence on the quench time with a tunable exponent \cite{Diptiman08,Barankov08,GomezRuiz19}. This feature is also found in the finite-time cooling of an Ising ferromagnet under Glauber dynamics \cite{Krapivsky10,Jeong20,priyanka2020slow} and we next study its effect on the distribution of kinks and the cumulant scaling.
To this end, we consider the algebraic cooling schedule
\begin{equation}
1-\gamma(t) \approx A\left( 1- \frac{t}{\tau_{Q}}\right)^{\alpha},
\end{equation} 
parameterized by $\alpha$.
The analogous form of equation (\ref{linearintegral}) is thus
\begin{equation}
\label{algWn}
W_{n}(\tau_{Q})=n\int_{0}^{2\tau_{Q}(1-\frac{A}{1+\alpha})}\mathrm{d}\eta e^{-\eta-\frac{A}{1+\alpha}\frac{\eta^{1+\alpha}}{(2\tau_{Q})^{\alpha}}}\frac{I_{n}(\eta)}{\eta},
\end{equation}
where now
\begin{equation}
\eta= 2\int_{\tau}^{\tau_{Q}}\mathrm{d}t\gamma(t) \approx 2(\tau_{Q}-\tau)-2\frac{A\tau_{Q}}{\alpha+1}\left(1-\frac{\tau}{\tau_{Q}}\right).
\end{equation} 
Again asymptotically approximating (\ref{algWn}) in the limit of large $\tau_{Q}$ we find the general expression
\begin{align}
\label{corralg}
\begin{split}
W_{n} = 1 &- n\sqrt{\frac{2c(\alpha)}{\pi}}\frac{1}{(2\tau_{Q})^{\frac{\alpha}{2(1+\alpha)}}}\left[\Gamma\left(\frac{3}{4}\right) \right.\\ &\left.-\frac{c(\alpha)(4n^2-1)}{24(2\tau_{Q})^{\frac{\alpha}{1+\alpha}}}\Gamma\left(\frac{1}{4}\right)+\mathcal{O}\left(\frac{1}{\tau_{Q}^{\frac{2\alpha}{1+\alpha}}}\right)\right],
\end{split}
\end{align}
where $c(\alpha)=(\frac{A}{1+\alpha})^{\frac{1}{1+\alpha}}$. The expression for $\kappa_{1}$ is thus
\begin{equation}
\label{kappa1alg}
\kappa_{1} = \frac{N}{2}\sqrt{\frac{2c(\alpha)}{\pi}}\frac{\Gamma\left(\frac{3}{4}\right) }{(2\tau_{Q})^{\frac{\alpha}{2(1+\alpha)}}} + \frac{N}{16}\sqrt{\frac{2}{\pi}}\frac{c(\alpha)^{\frac{3}{2}}\Gamma\left(\frac{1}{4}\right)}{(2\tau_{Q})^{\frac{3\alpha}{2(1+\alpha)}}} + \dots.
\end{equation} 
Comparing again leading and subleading amplitudes, we find
\begin{equation}
\label{tau1alg}
\tau_{Q}^{(1)}\gg\frac{1}{2}\left(\frac{A}{1+\alpha}\right)^{\frac{1}{\alpha}}\left[\frac{1}{8}\frac{\Gamma(\frac{1}{4})}{\Gamma(\frac{3}{4})}\right]^{1+\frac{1}{\alpha}}.
\end{equation}
A straightforward exercise verifies that (\ref{corralg}), (\ref{kappa1alg}) and (\ref{tau1alg}) coincide with the respective linear schedule expressions (\ref{corrlin}), (\ref{kappa1lin}) and (\ref{tau1lin}) in the special case $\alpha=A=1$. To leading order in  $1/\tau_Q$,
\begin{equation}
W_{n}\approx 1-n\sqrt{\frac{2c(\alpha)}{\pi}}\frac{\Gamma(\frac{3}{4})}{(2\tau_{Q})^{\frac{\alpha}{2(1+\alpha)}}},
\end{equation}
where $c(\alpha)=(\frac{A}{1+\alpha})^{\frac{1}{1+\alpha}}$. Thus, the expressions analogous to (\ref{kappaqlin}) are, keeping only the leading order in  $1/\tau_Q$,
\begin{equation}
\label{eq:kappa_nonlinear}
\kappa_{1}=\kappa_{2}=\kappa_{3}=N\sqrt{\frac{c(\alpha)}{2\pi}}\frac{\Gamma(\frac{3}{4})}{(2\tau_{Q})^{\frac{\alpha}{2(1+\alpha)}}}.
\end{equation}
Cumulants of the kink distribution thus exhibit a power-law scaling with the quench time. In particular, the power-law exponent $\frac{\alpha}{2(1+\alpha)}$  increases within its range $[0,1/2]$ as the parameter $\alpha$  of the cooling protocol is increased.

\begin{figure*}[t]
	\includegraphics[width=0.32\linewidth]{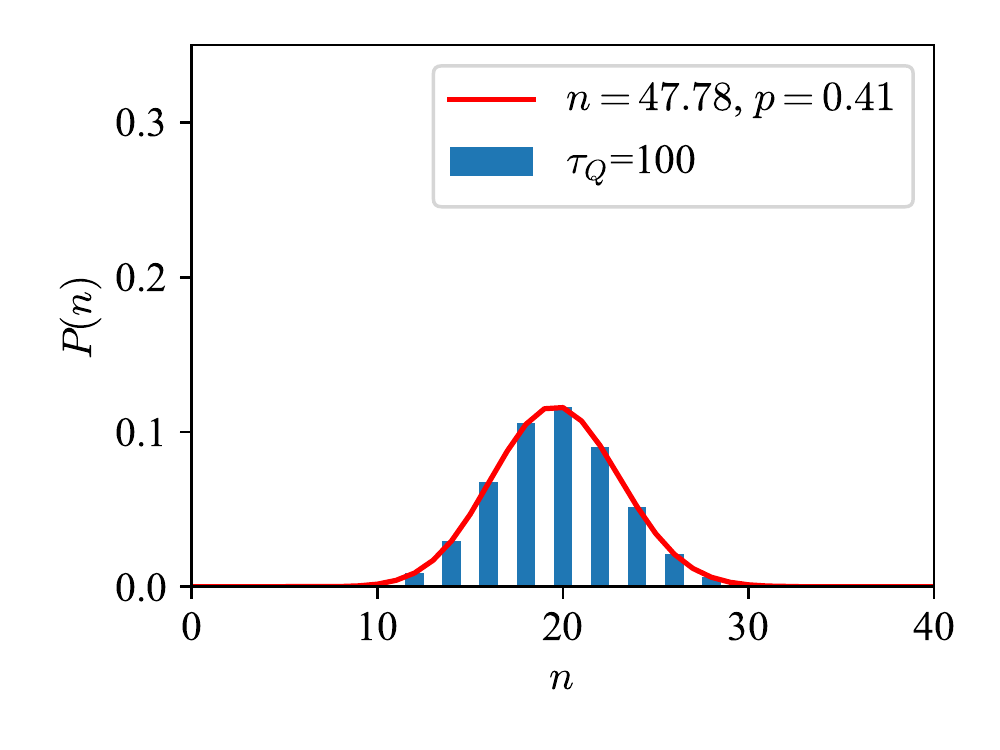}
	\includegraphics[width=0.32\linewidth]{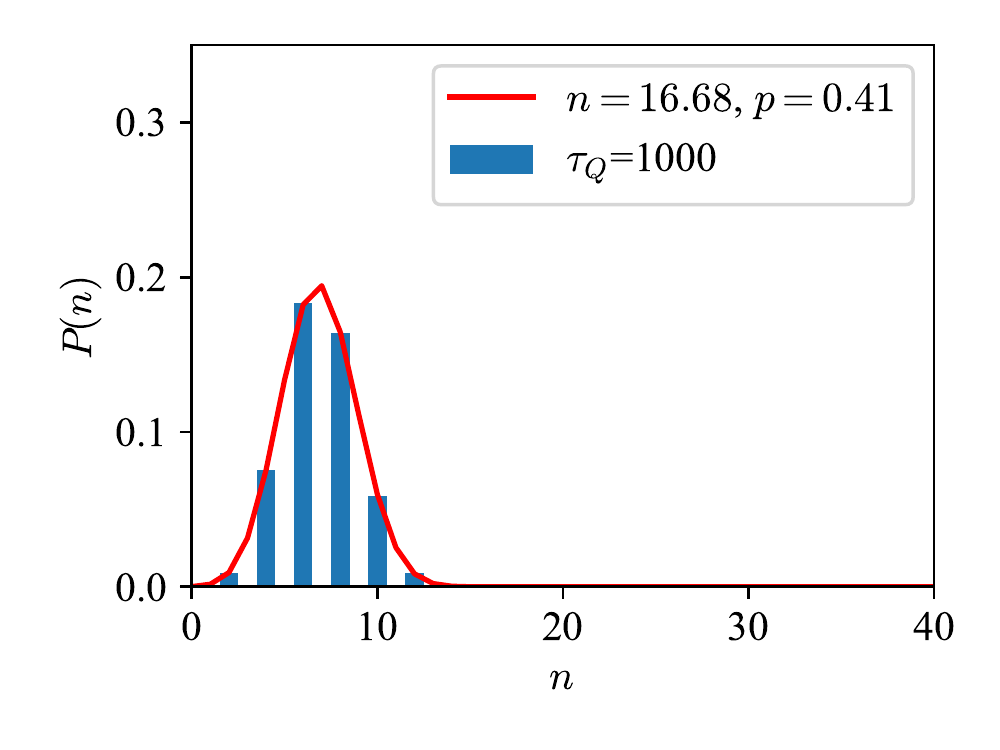}
	\includegraphics[width=0.32\linewidth]{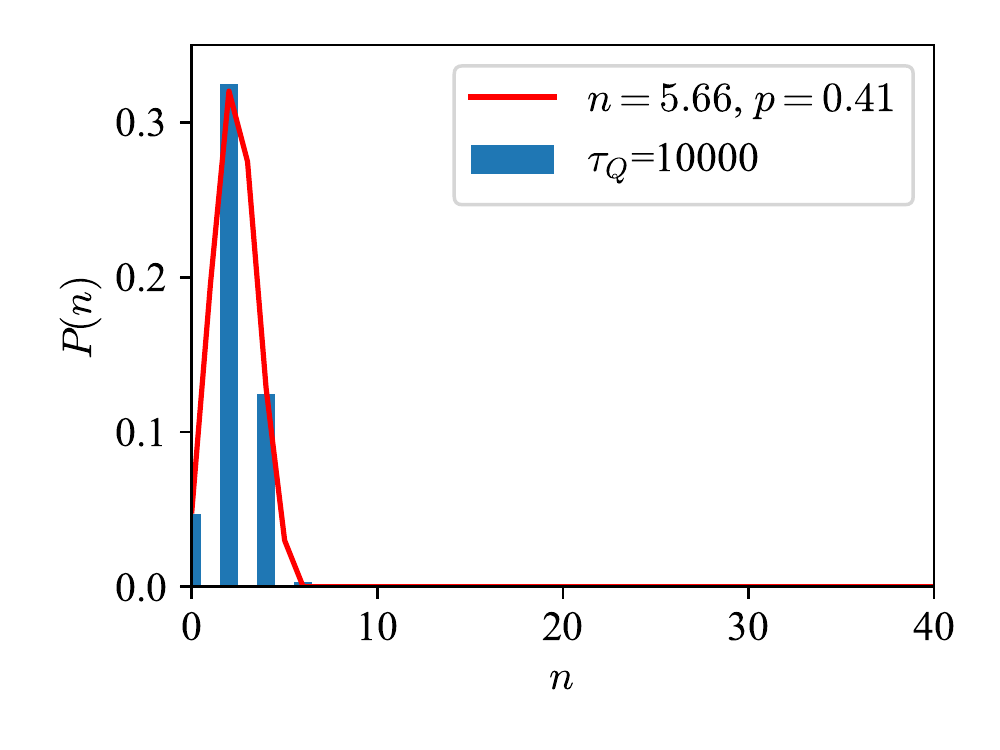}
	\caption{Kink number distribution and corresponding binomial distributions at different cooling rate for exponential cooling schedules.The system size is $N=500$ and the kink number distributions are obtain from $M=500000$ independent Glauber dynamics simulations. At the onset of adiabatic dynamics the distribution becomes asymmetric.}
		\label{fig:kink-dist-exp}
\end{figure*}

\begin{figure}[h]
	\includegraphics[width=\linewidth]{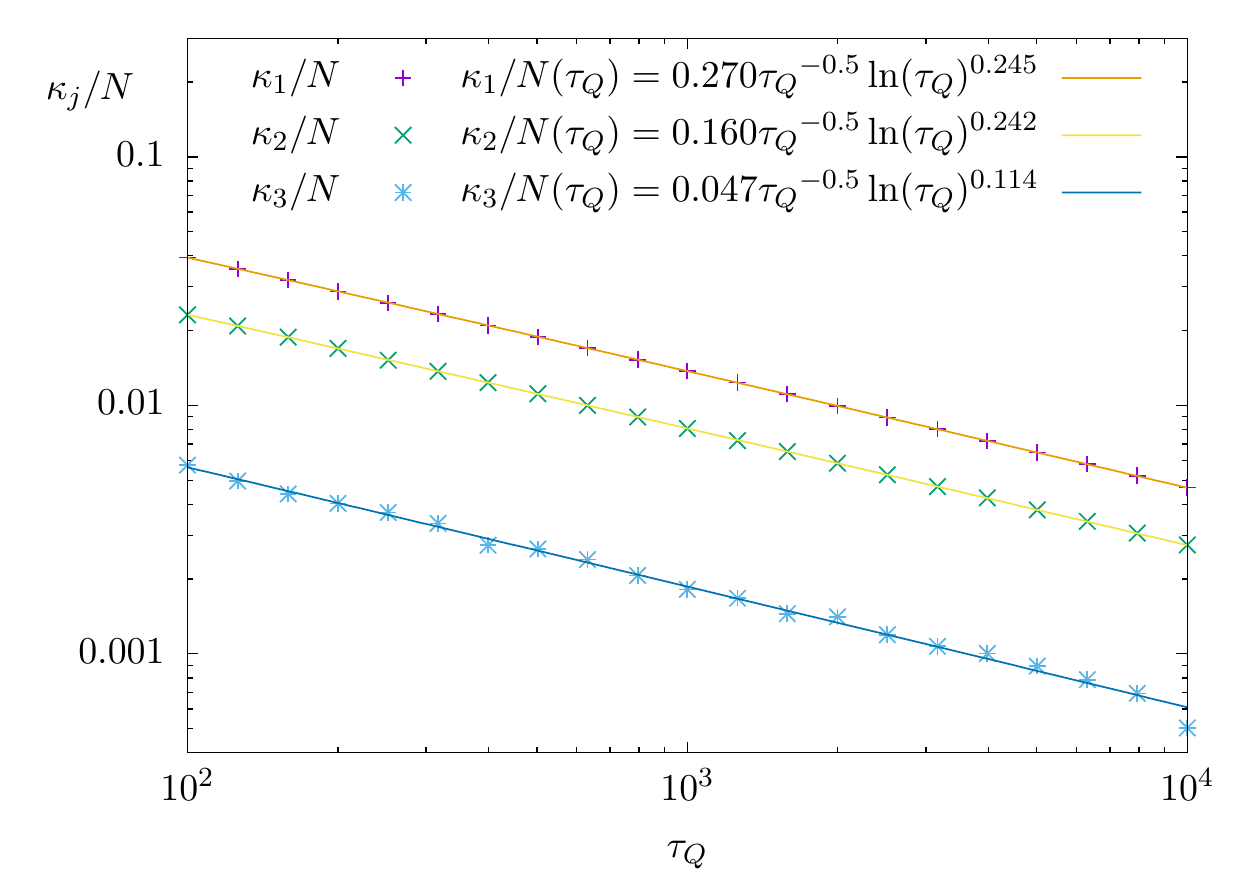}
	\includegraphics[width=\linewidth]{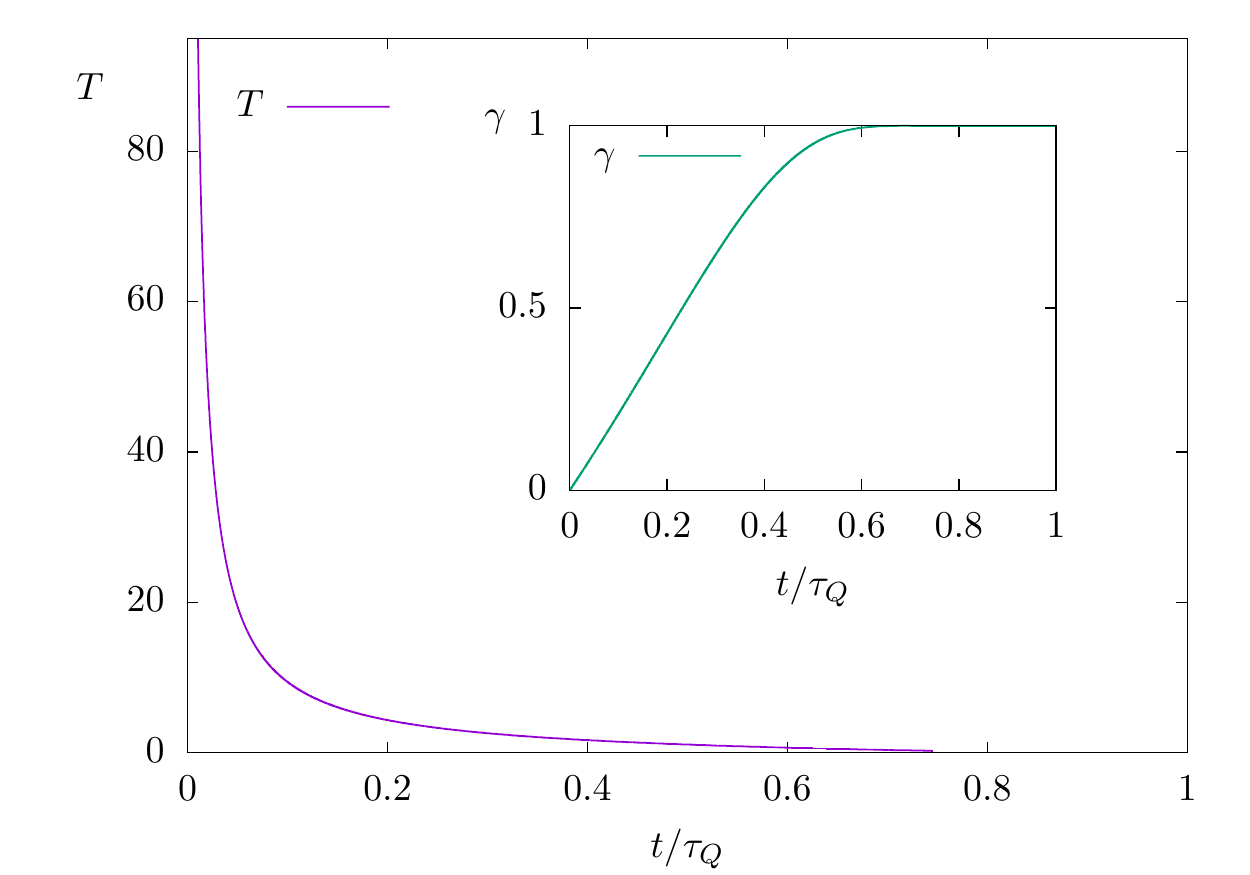}
	\caption{First three cumulants scaling of exponential cooling schedules (top panel) and the cooling schedule (bottom panel). The system size is $N= 500$. For all three cumulants, each data point is obtained by averaging over $M = 500000$ independent Glauber dynamics simulations.}
		\label{fig:cumulants_exp}
\end{figure}

Minimizing $\kappa_j$ ($j=1,2,3$) with respect to $\alpha$ we find that the optimal value of $\alpha$
\begin{equation}
\alpha_*=2Ae\tau_{Q}-1,
\end{equation}
which yields the minimum value of the cumulants
\begin{equation}
\label{eq:ONL}
\kappa_j(\alpha_*)
=\frac{N\Gamma(\frac{3}{4})}{\sqrt{2\pi}}\left[\frac{1}{\sqrt{2\tau_Q}}-\frac{1}{2Ae}\frac{1}{(2\tau_Q)^{3/2}}+\mathcal{O}(\tau^{-5/2})\right].
\end{equation}
Comparing the mean number of kinks resulting from this  optimized nonlinear schedule and the linear case in Eq. (\ref{kappaqlin}), we find that the latter leads to an enhanced suppression by a factor 
\beqa
\frac{\kappa_1(\alpha_*)}{\kappa_1^{\rm lin}}=\frac{\kappa_j(\alpha_*)}{\kappa_j^{\rm lin}}=\frac{1}{\tau_Q^{1/4}}.
\eeqa
Said differently, for a given cooling time $\tau_Q$ it is possible to reduce the mean number of kinks with respect to the linear schedule by using an algebraic schedule. 
This finding is reminiscent of the suppression of the mean number of excitations in the quantum dynamics of isolated critical systems, in which the dynamics is unitary and thus preserves entropy along the evolution \cite{Diptiman08,Barankov08,GomezRuiz19}. Here, we further note that the same conclusion applies to higher-order cumulants.
However, as discussed in Sec. \ref{SecSudden}, a sudden quench outperforms these schedules.

Fig.~\ref{fig:kink-dist-nonlinear} shows the distribution function of kink-number $P(n)$ obtained from  Glauber dynamics simulations for three different annealing rates. Again, the kink-statistics is well described by binomial distribution parametrized by $n=\kappa_1p$, $p=1-\kappa_2/\kappa_1$.
For the algebraic cooling schedule with $\alpha = 2$ and $A = 1$, dependence of the first three cumulants on annealing rate again exhibits a power-law relation, as shown in Fig.~\ref{fig:cumulants_nonlinear}. The fitted exponents for the three cumulants are $0.331\pm0.001$, $0.329\pm0.001$, and $0.336\pm0.011$, which are close to the value $\frac{\alpha}{2(1+\alpha)} = \frac{1}{3}$ predicted in Eq.~(\ref{eq:kappa_nonlinear}).

\subsection{Exponential Quench}

Both linear and algebraic cooling schedules lead to a power-law scaling of the cumulants of the kink distribution.
We next consider an exponential quench in the form suggested by Krapivsky \cite{Krapivsky10} 
\begin{equation}
1-\gamma(t) \approx B\exp \left\{-\frac{b}{\left(1-\frac{t}{\tau_{Q}}\right)^{\beta}} \right\}.
\end{equation}
Here, $b, \beta >0$ are positive real coefficients and $B=\exp(b)$ is a normalization factor ensuring $\gamma(0)=0$ and $\gamma(\tau_{Q})=1$. Making the substitution $\eta = 2h(\tau_{Q},\tau)$, we find the rather cumbersome integral expression
\begin{align}
\begin{split}
\label{expint}
&n\int_{0}^{2\tau_{Q}\left(1-\frac{Bb^{1/\beta}}{\beta}\Gamma\left(-1/\beta,b\right)\right)}\mathrm{d}\eta\exp\Bigg\{-\eta  \\ &  - \frac{B}{b\beta}\left(\frac{\eta}{2\tau_{Q}}\right)^{\beta}\eta\exp\left[ -b\left(\frac{\eta}{2\tau_{Q}}\right)^{-\beta}\right] \Bigg\}\frac{I_{n}(\eta)}{\eta}.
\end{split}
\end{align}
where $\Gamma(a,b)$ denotes the (upper) incomplete gamma function. Asymptotic solution of the integral (\ref{expint}) then leads to the result for $W_{n}$
\begin{equation}
W_{n}=1- n \frac{1}{\sqrt{\pi\tau_{Q}}}\left(\frac{\ln(\tau_{Q})}{b}\right)^{\frac{1}{2\beta}},
\end{equation}
giving
\begin{equation}
\label{eq:kappa-exp}
\kappa_{1}=\kappa_{2}=\kappa_{3}=\frac{N}{2\sqrt{\pi\tau_{Q}}}\left(\frac{\ln(\tau_{Q})}{b}\right)^{\frac{1}{2\beta}}.
\end{equation}

Fig.~\ref{fig:kink-dist-exp} shows the numerical distribution function of kink-number $P(n)$ for three different annealing rates. The kink-statistics is also well described by binomial distribution parametrized by $n=\kappa_1p$, $p=1-\kappa_2/\kappa_1$. 
Fig.~\ref{fig:cumulants_exp} shows the first three cumulants as functions of the annealing rate using parameters $\beta = 2$, $b = 1$ and $B = 1$ for the exponential cooling schedule. Consistent with the analytical prediction Eq.~(\ref{eq:kappa-exp}), we find all three cumulants exhibit a power-law dependence as a function of $\log(\tau_Q)$. 
However, contrary to a constant exponent $\frac{1}{2\beta} = 0.25$,  our best nonlinear least-square fit gives three different values $0.245\pm0.001, 0.241\pm0.002, 0.114\pm0.040$ for the exponents. In particular, the third cumulant is substantially different from the theoretical prediction. 
For $\beta=2$,  the $\log(\tau_Q)^{1/2\beta}$ term is changing very slowly in the $\tau_Q$ range of the simulation, making the cumulants only weakly dependent on $\log(\tau_Q)$; thus, the $\tau_Q$ dependence of cumulants is dominated by the $1/\sqrt x$ term in this $\tau_Q$ range. This makes the higher-order cumulant fitting in this cooling schedule more sensitive to numerical uncertainties. 

A remarkable feature of the  exponential schedule is  that its cooling efficiency surpasses that of the optimized nonlinear schedule. Indeed, taking the ratio of (\ref{eq:ONL}) over (\ref{eq:kappa-exp}) we find
\beqa
\frac{\kappa_j(\alpha_*)}{\kappa_{j}^{\rm exp}}=\Gamma\left(\frac{3}{4}\right)\left(\frac{\ln\tau_{Q}}{b}\right)^{-\frac{1}{2\beta}},
\eeqa
that is, the exponential schedule leads to a logarithmic suppression of the mean kink density with the quench time over the optimized nonlinear schedule.

\section{Thermal equilibrium at arbitrary temperature}

We briefly consider the equilibrium kink number distribution that will be relevant to the following sections devoted to sudden quenches and non-thermal behavior. We consider an arbitrary inverse temperature $\beta\geq 0$. In this case, it is known that the $k$-point correlator takes the form \cite{Baxter,Aliev98}
\beqa
\la \sigma_{i_1}\cdots \sigma_{i_k}\ra=z^{i_2-i_1+\cdots+i_k-i_{k-1}},
\eeqa
with
\beqa
z=\frac{1-\sqrt{1-\gamma^2}}{\gamma}=\tanh(\beta J).
\eeqa
It follows that at equilibrium two-point  correlator at distance $n$ equals
\beqa
W_n=z^n.
\eeqa
Using the expressions (\ref{kappa1}), (\ref{k2w})  and (\ref{k3w}) for $\kappa_j$ ($j=1,2,3$), one obtains
\beqa
\kappa_1&=&\frac{N}{2}(1-z),\\
\kappa_2&=&\frac{N}{4}(1-z^2),\\
\kappa_3&=&\frac{N}{4}z(1-z^2),\
\eeqa
We note that these expressions are equivalent to those of the binomial distribution $B(N,p)$
\beqa
\label{kappabinom}
\kappa_{1}&=&Np,\\
 \kappa_{2}&=&Np(1-p),\\ 
 \kappa_{3}&=&Np(1-p)(1-2p),
\eeqa
 with the kink formation probability
 \beqa
 \label{pthermal}
 p=\frac{1-z}{2}=\frac{1-\tanh(\beta J)}{2}.
 \eeqa
In the infinite temperature case, the distribution describes as well that of the quantum Ising chain \cite{Michal21a}.

\section{Fast and Sudden Quenches}\label{SecSudden}
We next consider the behavior of the system under a rapid quench and note that each cooling schedule yields in this limit
\begin{equation}
W_{n}(\tau_{Q})\approx n\int_{0}^{\tau_{Q}}\mathrm{d}\eta e^{-\eta}\frac{I_{n}(\eta)}{\eta}.
\end{equation}
Taking the series expansion of $I_{n}$ at $\eta \rightarrow 0$, we find that
\begin{equation}
\label{shortlimiteq}
W_{n}(\tau_{Q})=n\sum_{k=0}^{\infty}\frac{1}{\Gamma(k+n+1)k!}\frac{1}{2^{2k+n}}\int_{0}^{\tau_{Q}}\mathrm{d}\eta e^{-\eta}\eta^{2k+n-1}.
\end{equation}
The integral in (\ref{shortlimiteq}) is equal to the lower incomplete gamma function $\gamma(a,b)$, which gives a Taylor series, the leading factor of which is of the form $b^{a}\Gamma(a)e^{-b}$. Taking the leading factors of both summations, observing that $e^{-\tau_{Q}}\approx 1-\tau_{Q}$ for fast quenches and taking the minimal power in $\tau_{Q}$ yields
\begin{equation}
\label{shortq}
W_{n}(\tau_{Q})\approx \left(\frac{\tau_{Q}}{2}\right)^{n}.
\end{equation}
Substituting (\ref{shortq}) into the expressions for $\kappa_{1}$, $\kappa_{2}$ and $\kappa_{3}$, while taking the leading powers gives
\beqa
\label{kappafast}
\kappa_{1} &=& \frac{N}{2}\left(1-\frac{\tau_{Q}}{2}\right) , \\
\kappa_{2} &=& \frac{N}{4}\left(1-\frac{\tau_{Q}^{2}}{4}\right), \\
 \kappa_{3} &=& \frac{N\tau_{Q}}{8}.
\eeqa
The kink distribution upon completion of the quench in the limit of vanishing $\tau_Q$ is that of a ferromagnet at infinite temperature. 
According to Eq. (\ref{pthermal}), for $\beta=0$ the kink formation probability is $p=1/2$, as expected.
As a result, the cumulant values of a binomial distribution $B(N,1/2)$ in Eq. (\ref{kappabinom})
are recovered, i.e.,  $\kappa_{1} = \frac{N}{2}$, $\kappa_{2} =\frac{N}{4}$, $\kappa_3=0$.
We further notice that the values in Eqs. (\ref{kappafast}) also agree with those of the binomial distribution in (\ref{kappabinom}) when the kink formation probability reads
\beqa
p=\frac{1}{2}\left(1-\frac{\tau_Q}{2}\right),
\eeqa
which captures the leading correction away from the sudden limit due to the finite value of the quench time $\tau_Q$.

We note in passing that another interesting dynamical phenomenon related to sudden quench is domain coarsening~\cite{bray02coarsening}. It is generally believed that coarsening systems exhibit dynamical scaling, i.e., the typical domain size grows algebraically with time $L \sim t^{1/z}$, where $z$ is a dynamical exponent that is independent of microscopic details of the system. In the 1D Ising chain, the typical domain size is simply related to the average distance between kinks, hence $L \sim 1/\kappa_1$. The scaling hypothesis thus implies a power-law behavior for the first cumulant.  Interestingly, our extensive Glauber dynamics simulations show that all three cumulants follow a diffusive scaling law: $\kappa_j \sim t^{-1/2}$, corresponding to an exponent $z = 2$; see Appendix~\ref{sec:coarsening} for more details.  
As a caveat, it should be noted that the phenomenology of sudden thermal quenches differs from that of sudden quenches in quantum phase transitions in isolated spin chains. Indeed,  the sudden quench followed by an evolution time leads to a lower density of defects than the linear, nonlinear, and exponential schedules for a given total duration of the process.

\section{Non-thermal behavior}
One may wonder whether the non-equilibrium state resulting from the finite-time cooling of a ferromagnet is effectively thermal. To that end, one can compute the distance between an equilibrium thermal distribution of kinks $P_{\beta}(n)$ with inverse temperature $\beta$ as a free parameter and the numerically obtained distribution $P(n)$ for given $P(n)=P_{\tau_Q}(n)$. The proximity between the two distributions can be quantified by a distance. 
We consider the trace-norm distance
 \beqa
 D_{\rm TN}=\frac{1}{2}\sum_n\left|P_{\beta}(n)-P_{\tau_Q}(n)\right|.
 \eeqa
 Minimizing it with respect to  the free parameter $\beta$, 
 \beqa
 \min_{\beta} D_{\rm TN}=D_{\rm TN}^*
 \eeqa
one can identify the effective temperature $\beta^*$ that best approximates the non-equilibrium state with distance $D_{\rm TN}^*$.
The equilibrium distribution $P_{\beta}(n)$ is obtained from standard Monte Carlo simulation using Glauber dynamic spin update and the trace-norm distance is minimized using golden search method.
\begin{figure}[t]
	\includegraphics[width=\linewidth]{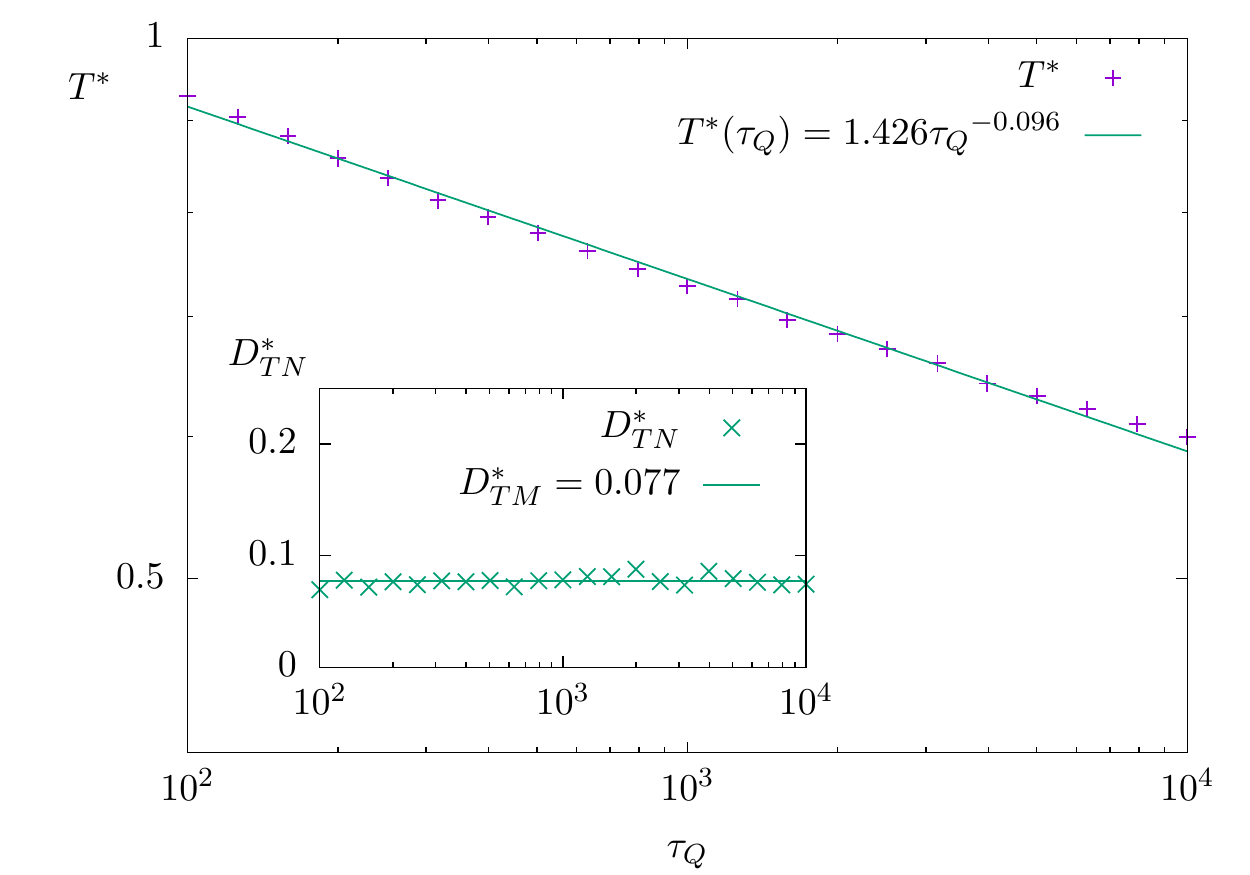}
	\caption{Effective temperature (main panel) and trace-norm distance (inset panel) for linear cooling schedule. 	\label{figureTeff}}
\end{figure}
For concreteness, we focus on the case of a linear quench protocol.
The numerical simulations for a chain of $L=500$ spins indicate non-thermal behavior in the final state for quench times $\tau_Q\in[10^2,10^4]$, see Fig. \ref{figureTeff}. For these parameters, the minimum trace norm distance remains in the interval $ D_{\rm TN}^*\in [0.07,0.09]$. The canonical Gibbs state that best approximates the final state is characterized by an inverse temperature that scales as a power law of the quench time with exponent $-0.096\pm0.001$.
We note that in a Gibbs state, the mean kink number  equals \cite{Xu19}
\beqa
\kappa_1=\frac{N}{1+e^{2\beta J}}.
\eeqa 
 Assuming $D_{\rm TN}=0$, and comparing this expression with the result for a linear quench (\ref{kappaqlin}) 
 suggests that the power-law scaling is effective and results from linearizing the logarithmic dependence (taking the Boltzmann constant $k_B=1$)
 \beqa
T^*(\tau_Q)=
\frac{2 J}{\log \left(\frac{2 \sqrt{\pi }  \sqrt{\tau_Q}}{ \Gamma
   \left(\frac{3}{4}\right)}-1\right)},
 \eeqa
 over the studied range of quench times. Naturally, the explicit dependence  of $T^*$ on $\tau_Q$ varies with the cooling schedule.

\section{Limit of numerical simulation}

The deviations between the analytical results and the numerical data observed in the histograms and cumulant scaling behavior come from the fact that the slow cooling limit $\tau_Q \rightarrow \infty$ and the thermodynamic limit of infinite system size, both considered in the analytical approach, are not accessible in the numerical simulation. 
In the low temperature, long-time limit, the topological defects are exponentially scarce, making finite-size effects significant in the slow cooling regime. 
An arbitrarily long cooling time will simply bring a finite system close to equilibrium and the non-equilibrium physics can not be fully captured. 
Therefore, to explore the non-equilibrium physics in the slowing cooling limit, the infinite size limit is also required, which is beyond reach in the numerical simulation.
The asymptotic behavior of the system in slow cooling limit can still be analyzed by observing the scaling of  the correlator $W_n$, which has a stronger dependence on the cooling rate. 
\begin{figure}[t]
	\label{linear cooling correlator}
	\includegraphics[width=\linewidth]{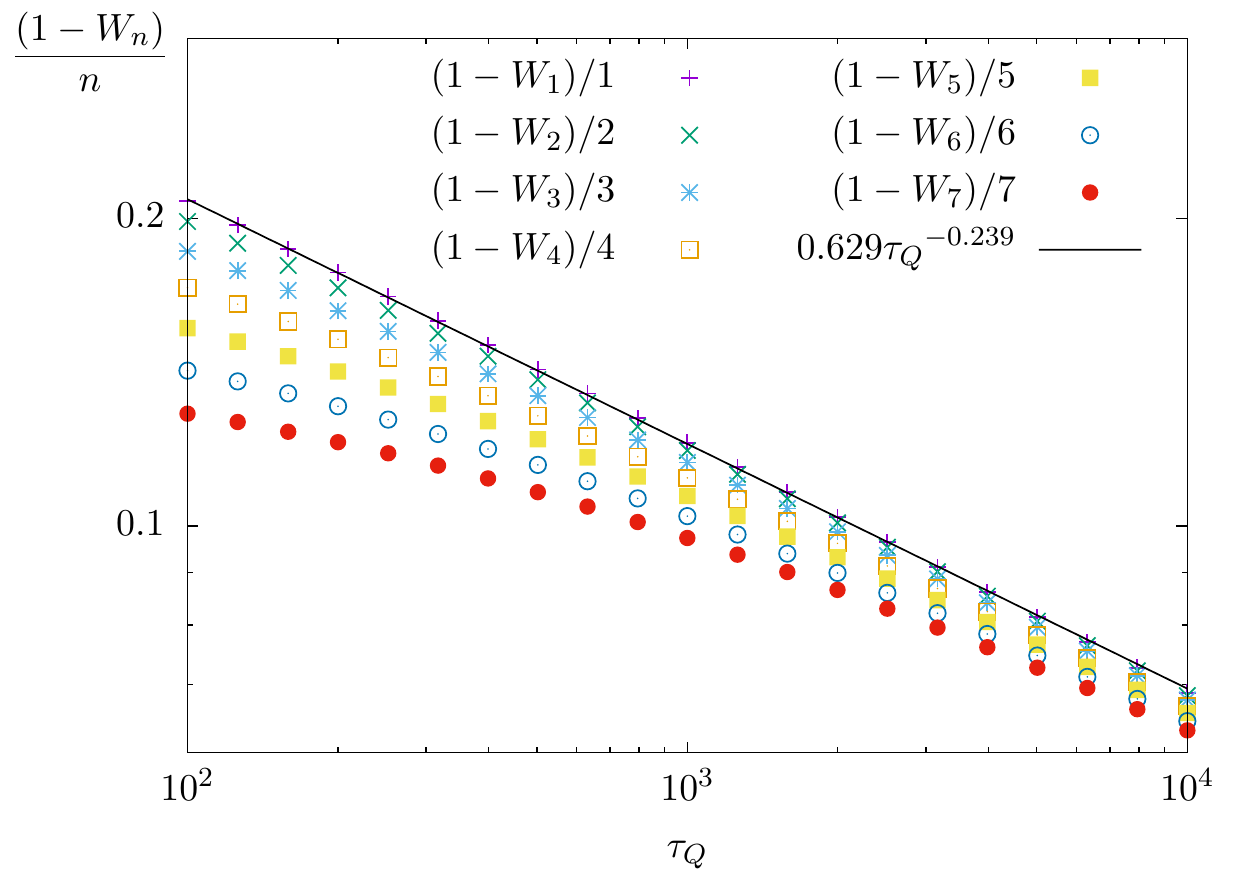}
	\caption{Scaling of correlator function $1-W_{n}$ for linear cooling schedule. In the limit of slow quench times the ratio $(1-W_{n})/n$ is independent of $n$ and governed by the leading term in Eq. (\ref{corrlin}).}
\end{figure}
It can be seen that $W_n$ converges to the predicted expression $1-nC'\tau_Q^{-\delta}$ in the long time limit.
This dependence will further lead to the cumulant behavior predicted by the analytical calculation. 
Note that in the range of quench times $\tau_Q\in[10^2,10^4]$, the system size does not have a significant impact on the results.

\section{Connection to the Kibble-Zurek Mechanism: Mean number of kinks}

KZM predicts the mean density of defects upon completion of a cooling schedule making use of equilibrium properties \cite{DZ14}.
We first recall the equilibrium correlation length of the one-dimensional Ising model is \cite{Plischke06}
\beqa
\xi=\frac{\xi_0}{|\log(\tanh(\beta J))|},
\eeqa
where $\xi_0$ is the lattice spacing. The one-dimensional ferromagnet thus differs from the standard setting 
in higher dimensions, where correlation length exhibits a power-law scaling as a function of the proximity to the critical point.
By contrast, the relaxation time under Glauber dynamics exhibits the conventional power-law divergence \cite{Krapivsky10,Jeong20}
\beqa
\tau=\frac{\tau_0}{|1-\gamma|},
\eeqa
where $\tau_0$ is a microscopic constant. As the critical point at $T=0$, the system exhibits critical slowing down and the dynamics can be expected to be nonadiabatic for any finite quench time.
For the sake of illustration, we focus on the linear cooling schedule 
\beqa
\gamma(t)=\frac{t}{\tau_Q}=\tanh(2\beta J).
\eeqa
KZM invokes the adiabatic impulse approximation, according to which the relaxation time is in an early stage small enough so that the system quickly adjusts to the instantaneous equilibrium configuration with $\gamma=\gamma(t)$. The growth of the relaxation time close to the critical point gives rise to the effective freezing of the order parameter of the system. 
KZM estimates the mean size of the domains (out-of-equilibrium correlation length) after cooling in finite time by the equilibrium value of the instantaneous correlation length at freezing, the so-called freeze-out time $\hat{t}$.
To estimate  the freeze-out  time $\hat{t}$, we match the instantaneous equilibrium relaxation time
to the time left until reaching the critical point $\tau_Q-t$, that is,
\beqa
\tau(t)= \tau_Q-t.
\eeqa
For the linear schedule, the solution is given by
\beqa
\hat{t}=\sqrt{\tau_0\tau_Q}.
\eeqa
By an analogous procedure, one can estimate the freeze-out-time for other schedules such as the algebraic and the exponential one.
Using the relation between the correlation length and the relaxation time
\beqa
\xi=\xi_0\left(\frac{\tau}{\tau_0}\right)^{\frac{1}{z}},
\eeqa
KZM predicts the mean domain size after cooling to be given by
\beqa
\hat{\xi}=\xi(\hat{t})=\xi_0\left(\frac{\tau_Q}{\tau_0}\right)^{\frac{1}{2z}},
\eeqa
which for $z=2$ yields the  power-law scaling
\beqa
\la \hat{\mathcal{N}}\ra=\frac{N}{\hat{\xi}}\propto\tau_Q^{-1/4}.
\eeqa
The accuracy of the KZM in accounting for the finite-time cooling of the Glauber dynamics has been discussed in
\cite{Krapivsky10,Jeong20}. We next focus on physics beyond KZM associated with the kink number statistics.

\begin{figure*}
	\includegraphics[width=0.7\linewidth]{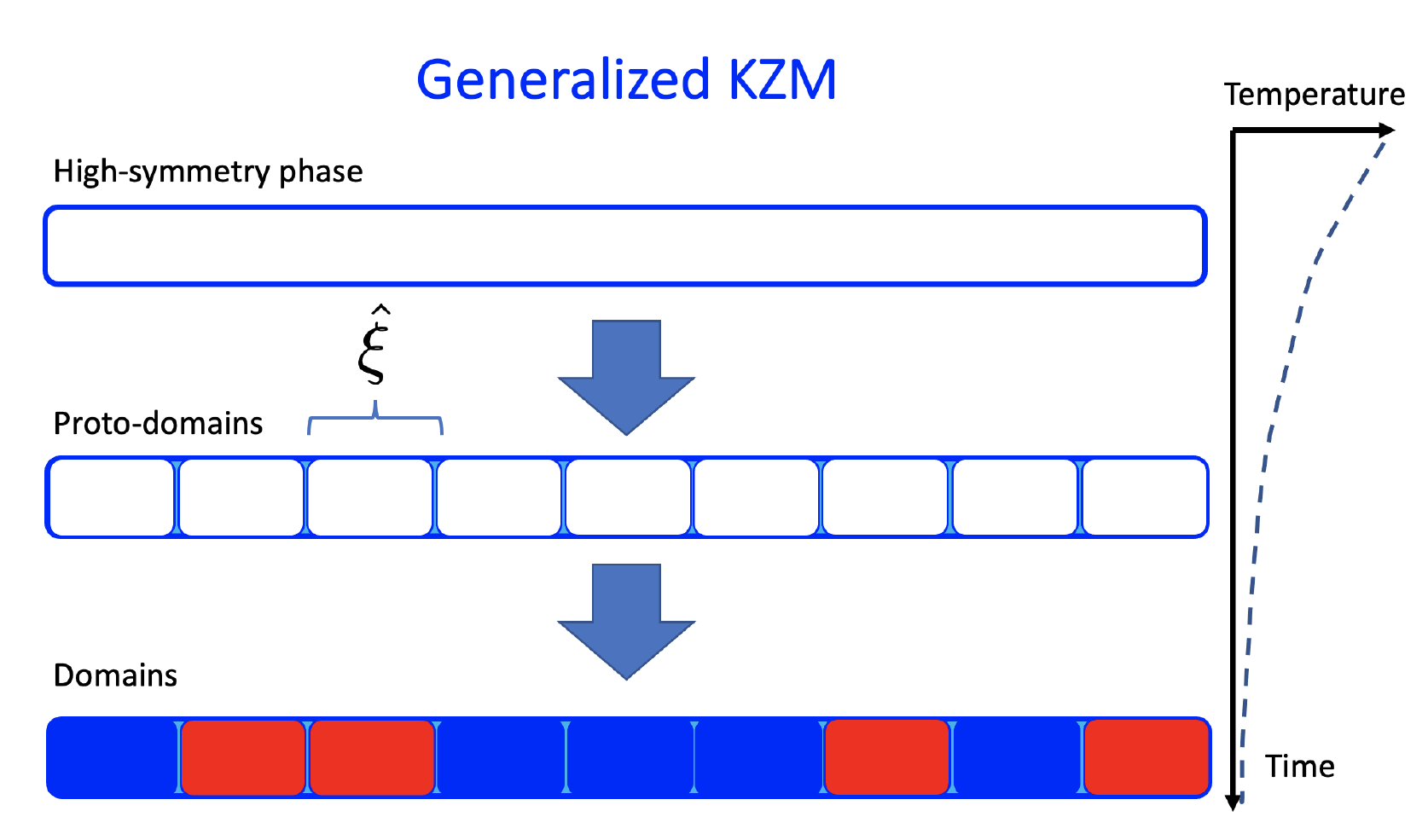}
	\caption{Schematic representation of the annealing dynamics according to the generalized Kibble-Zurek mechanism. 
	In the course of the annealing, a system of size $L$ is partitioned into proto-domains of the KZM length scale $\hat{\xi}$, which scales as a power-law with the quench time. At the interface between adjacent proto-domains, kinks are spontaneously formed with probability $p$. There are $\mathrm{N}_b=L/\hat{\xi}$ interfaces. Assuming events of kink formation to be uncorrelated at different locations yields a binomial distribution for the kink number distribution $P(n)\sim B(n,\mathrm{N}_b,p)$, in which all cumulants scale as $\hat{\xi}^{-1}$. Domains with opposite spin configurations are represented in blue and red color and are separated by kinks.}
		\label{fig:GenKZM}
\end{figure*}
\section{Beyond the Kibble-Zurek Mechanism: Kink number statistics}\label{Sec:GenKZM}

A growing body of results \cite{Cincio07,delcampo18,Cui2019,GRMdC19,Bando20,delcampo2021universal}  suggests that the signatures of universality govern the kink number distribution and not only its mean value. To appreciate this, in a classical setting, it suffices to assume that the formation of kinks at different locations is described by independent stochastic events \cite{GRMdC19}.

\begin{figure}
	\includegraphics[width=\linewidth]{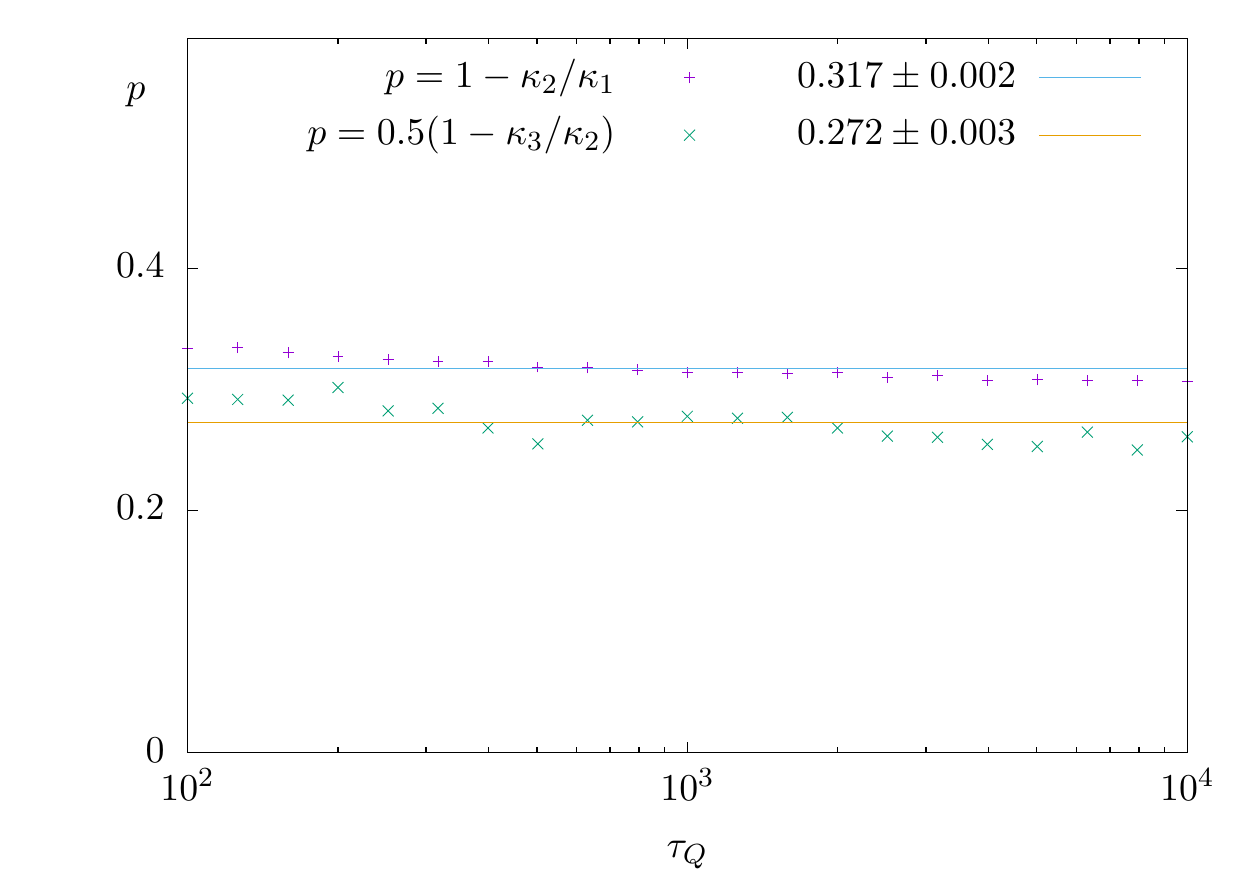}
	\caption{Value of the kink formation probability extracted from the cumulant ration. According to the generalized KZM, this value is independent of the quench time. The estimate of $p$ depends on the cumulant ratio considered. \label{FigPGenKZM}}
\end{figure}

The key tenet of KZM is that the cooling dynamics sets the average length scale of domains to be given by the equilibrium correlation length evaluated at the freeze-out time $\hat{\xi}$.
By contrast, to generalize KZM we consider that the effect of the cooling is to partition a  system of size $L=N\xi_0$ into ``proto-domains'' of the same length scale $\hat{\xi}$ over which the order parameter stabilizes. At the boundary between adjacent domains, kinks form with a given probability $p$. Conversely, with probability $(1-p)$ no kink is formed and the two adjacent proto-domains coalesce to form a larger domain. 
The number of  boundaries between proto-domains determines the number of stochastic events  for kink formation set by (the floor of)
\beqa
\mathrm{N}_b=\frac{L}{\hat{\xi}}=N\left(\frac{\tau_0}{\tau_Q}\right)^{\frac{1}{2z}},
\eeqa
where the second-equality holds for the Ising ferromagnet.
Assuming kink formation events at different locations to be uncorrelated leads to a kink number distribution associated with $\mathrm{N}_b$ independent and discrete random Bernoulli variables. 
Upon assuming the success probability $p$ to be the same at different locations, the distribution takes the binomial form
\beqa
P(n)= B(\mathrm{N}_b,p)= \begin{pmatrix}
\mathrm{N}_b  \\
n
\end{pmatrix}  p^n\,(1-p)^{\mathrm{N}_b-n},
\eeqa
see Figure \ref{fig:GenKZM}.
In one spatial dimension,
\beqa
\label{meankzm}
\kappa_1=\la n\ra=p \mathrm{N}_b=pN\left(\frac{\tau_0}{\tau_Q}\right)^{\frac{\nu}{1+z\nu}}.
\eeqa
Similarly, higher order cumulants of the binomial distribution read
\beqa
\kappa_2&=&(1-p) \kappa_1,\\
 \kappa_3&=&(1-2p)\kappa_2,\\ 
 \kappa_{q+1}&=&p(1-p)\frac{d\kappa_q}{dp}.
\eeqa
As a result, all cumulants are predicted to follow the same universal-power-law scaling predicted by the mean number of kinks, in agreement with the numerical simulations and analytical calculations we have reported.
In the binomial distribution, cumulant ratios are determined by the kink formation probability $p$, independent of the quench time.
Accordingly,  the numerical values of the cumulant ratios found in Fig. \ref{fig:cumulants_linear} are expected to be consistent with a well-defined probability for kink formation, and thus, with the binomial distribution. According to the generalized KZM this probability is independent of the quench time, as shown in Figure \ref{FigPGenKZM}.
From $\kappa_2/\kappa_1=1-p$, one finds $p=0.32$. Using any of the ratios involving the third cumulant, $\kappa_3/\kappa_2$, one finds the close value $p=0.27$, though we recall that the power-law scaling of $\kappa_3$ deviates from the KZM prediction. 
These values are comparable to those observed  in other one-dimensional systems, such as the overdamped Ginzburg-Landau model ($p\approx 0.42$) and the transverse-field quantum Ising model, well in isolation ($p=0.41$) \cite{Cincio07,delcampo18,Cui2019}, or coupled to a bath ($p\approx0.37-0.39$) \cite{Bando20},

\section{Discussion and conclusion}

We have analyzed the kink number distribution of an Ising chain thermally annealed under Glauber dynamics.
While generally it is well described by a binomial distribution, in the limit of slow annealing kink statistics becomes Poissonian. We have explicitly computed the two-point function and used it to derive the low-order cumulants of the distribution. 
Specifically, the mean number of kinks, the variance, and the third centered moment are identical and given by a power-law with the quench time in the limit of slow cooling. 

 The one-dimensional Ising model does not exhibit a phase transition and in the absence of a magnetic field becomes degenerate only at zero temperature. The annealing schedules we have reported involve positive temperatures, approaching only degeneracy at infinite time. As a result, the annealing of the ferromagnet does not involve the crossing of the critical point. 
 The situation is similar to that in recent experiments with colloidal monolayers that probe only ``half of the transition'' \cite{Keim15,delcampo15}. In principle, such a scenario does not preclude the appearance of KZM scaling \cite{DZ06,Chandran12,BD18}, although under slow cooling, the dynamics is inextricably woven with coarsening \cite{Prados97,Krapivsky10,Biroli10,Jeong20,priyanka2020slow}. 
 
The Ising ferromagnet in one spatial dimension does not exhibit a power-law divergence of the equilibrium correlation length. As a result, the correlation length critical exponent $\nu$ is not defined. This precludes the application of the KZM in its original form \cite{DZ14}.  However,  the appearance of power-law behavior can be established using the adiabatic-impulse approximation \cite{Krapivsky10,Jeong20,priyanka2020slow}, a core tenet of KZM.

Focusing on kink number fluctuations, we have characterized the full kink number distribution that exhibits signatures of universality as predicted by the generalized KZM \cite{GRMdC19}.
The dependence of the cumulants of the kink number distribution on the annealing time varies with the schedule. When the temperature is a linear function of time, all cumulants are shown to scale with a power-law of the quench time. When the annealing schedule involves a polynomial variation of the temperature with time,  a modified power law is observed. This generalizes to arbitrary cumulants the scaling prediction for the mean number of defects resulting from the nonlinear passage across a critical point \cite{Diptiman08,Barankov08,Nikoghosyan16,GRMdC19}. 
We have found corrections to the power-law behavior of cumulants when the temperature decays exponentially as a function of time, see as well \cite{Krapivsky10,Jeong20,priyanka2020slow} for the mean number. At variance with previous studies exploring  Berezinskii–Kosterlitz–Thouless phase transition \cite{Jeli11,Dziarmaga14} and holographic systems \cite{Chesler15},  the logarithmic corrections to KZM scaling that we have reported stem directly from the annealing schedule. 

We have further analyzed the dependence of the cooling efficiency for a given quench time as a function of the choice of the cooling schedule.  Nonlinear quenches are shown to reduce the residual density of kinks below the value obtained under a linear quench. This result is consistent with previous findings on nonlinear quenches \cite{Diptiman08,Barankov08,Krapivsky10,Nikoghosyan16,GRMdC19,Jeong20,priyanka2020slow}. The nonlinearity can be optimized to maximize this suppression as suggested in \cite{Barankov08}. An exponential cooling protocol proves even more efficient than the optimal nonlinear quenches in suppressing kink formation. Yet, a sudden quench to zero temperature with subsequent evolution for the same total time surpasses all to this end. The cooling scenario thus exhibits a different phenomenology from that observed in quantum phase transitions, where sudden quenches enhance defect formation over finite-time protocols.

In the opposite limit of sudden and nearly sudden quenches the cumulant values are those of a binomial distribution. As a result, the shape of the distribution varies from binomial to Poissonian as the cooling rate is decreased. Yet, even in the scaling regime for slow quenches, we have shown that in the regime of generalized scaling behavior the final state is nonthermal, by establishing the trace distance between the resulting kink-number distribution upon completion of the quench and the corresponding one for a canonical Gibbs state. The thermal state that best approximates the final state exhibits an effective temperature that scales as an inverse power-law of the quench time.

We hope that the current findings motivate new studies of the finite-time annealing dynamics of a ferromagnet beyond the KZM.  A natural generalization involves the inclusion of disorder, which is known to turn the power-law scaling on the quench time under a linear schedule into a logarithmic dependence \cite{Dziarmaga06,Caneva07,Suzuki09,Suzuki11}.  An analogous description may be invoked in one-spatial dimension \cite{Aliev00} and the full counting statistics, as well as the role of the schedule, remain unexplored in this context.  Similarly, one can envision studies in higher spatial dimensions, as well as with continuum and gauge symmetries \cite{Creutz85}.

We close by pointing out the relevance of the cooling dynamics of classical spin models in the benchmarking of quantum annealers and quantum simulators.
By embedding Ising models in quantum annealing devices, tests of the KZM   have been used to benchmark their performance \cite{Gardas18,Weinberg19,Bando20} and a study of the kink statistics can provide a stringent test, helping to elucidate the kind of dynamics emulated in these devices \cite{Bando20}. In addition, we note that in other setups the kink number distribution can be directly measured making use of single-qubit interferometry, whereby an auxiliary qubit is used to probe the state of the Ising ferromagnet \cite{Xu19}.

\section*{Acknowledgements}
It is a pleasure to thank Mikhail A. Aliev, Michał Białończyk, Leticia Cugliandolo and Fernando J. G\'omez-Ruiz for useful discussions. The authors further thank  Hidetoshi Nishimori for feedback on the manuscript, and  Alberto Carta for helpful suggestions leading to the eventual solution of the integral (\ref{linearintegral}). This work is supported by the Spanish  Ministerio de Ciencia e Innovaci\'on (PID2019-109007GA-I00). JM further thanks the hospitality and support of the DIPC during the early stages of the project.

\newpage
\begin{appendix}
\begin{widetext}

\section{Computation of the second and third cumulants}\label{appk23}
For convenience we introduce the notation
\begin{equation}
\partial_{i_{1},...,i_{n}}^{n} \equiv  \frac{\partial^{n}}{\partial \eta_{i_{1}},...,\partial \eta_{i_{n}}}.
\end{equation}
Using the cumulant generating function, the variance of the kink number is given by
\beqa
\label{kappa2}
\kappa_{2} &=& - \frac{1}{4}\left[\sum_{n,m}(\partial^{4}_{m+1,m,n+1,n}\Psi)|_{\{ \eta_{i}\}=0}  -\left(\sum_{n}(\partial^{2}_{n+1,n}\Psi)|_{\{ \eta_{i}\}=0}\right)^{2}\right] \nonumber \\
&=& -\frac{1}{4}\left[\sum_{n,m}\langle \sigma_{N} \sigma_{n+1}  \sigma_{m} \sigma_{m+1} \rangle - \left(\sum_{n}\langle \sigma_{N} \sigma_{n+1}^{z}\rangle\right)^{2}\right] \nonumber \\
&=&\frac{1}{4}\left[4 \langle \hat{\mathcal{N}}^{2} \rangle + N^{2} - 4N\langle \hat{\mathcal{N}} \rangle - \left(N^{2} + 4 \langle \hat{\mathcal{N}} \rangle^{2} - 4N \langle \hat{\mathcal{N}} \rangle \right)\right] 
\nonumber\\
&=& \langle \hat{\mathcal{N}}^{2} \rangle - \langle \hat{\mathcal{N}} \rangle ^{2}.
\eeqa 
The third cumulant equals the third centered-moment and is given by
\beqa
\label{kappa3}
&  \kappa_{3}&  = -\frac{1}{8}\Bigg[\sum_{n,m,l}(\partial^{6}_{l+1,l,m+1,m,n+1,n}\Psi)|_{\{ \eta_{i}\}=0}\nonumber - 3\left(\sum_{n}(\partial^{2}_{n+1,n}\Psi)|_{\{ \eta_{i}\}=0}\right)\left(\sum_{n}(\partial^{2}_{m+1,m,n+1,n}\Psi)|_{\{ \eta_{i}\}=0}\right) \nonumber\\ & & + 2 \left(\sum_{n}(\partial^{2}_{n+1,n}\Psi)|_{\{ \eta_{i}\}=0}\right)^{3}\Bigg] \nonumber\\ 
&  =& \frac{1}{8}\Big[ -N^{3}-12N\langle \hat{\mathcal{N}}^{2}\rangle + 6N^{2} \langle\hat{\mathcal{N}}\rangle \nonumber+3\left(4N\langle\hat{\mathcal{N}}^{2}\rangle+N^{3}-6N^{2}\langle\hat{\mathcal{N}}\rangle\right) -2\left(N^{3}-8\langle\hat{\mathcal{N}}\rangle^{3}-6N^{2}\langle\hat{\mathcal{N}}\rangle +12N\langle\hat{\mathcal{N}}\rangle^{2}\right)\Big]\nonumber\\
&  =& \langle\hat{\mathcal{N}}^{3}\rangle-3 \langle\hat{\mathcal{N}}\rangle \langle\hat{\mathcal{N}}^{2}\rangle + 2\langle\hat{\mathcal{N}}\rangle^{3}.\nonumber\\
\eeqa
Equations (\ref{kappa1}),  (\ref{kappa2}) and  (\ref{kappa3}) thus lead to the well-known results for the cumulants in terms of moments of the distribution and show the consistency of using the logarithm of (\ref{gfunc}) as the cumulant generating function.

\subsection{Explicit Calculation of $\kappa_{2}$}
We begin with the expression (\ref{kappa2}), and substitute in the explicit correlators $W_{n}$ as derived previously from $\Psi$. Lastly, we use translational invariance of the system to dispense with the index $m$ to find a condensed form of the expression.
\beqa
\label{xxx}
\kappa_{2} &=& - \frac{1}{4}\bigg[\sum_{n,m}(\partial^{4}_{m+1,m,n+1,n}\Psi)|_{\{ \eta_{i}\}=0} - (\sum_{n}(\partial^{2}_{n+1,n}\Psi)|_{\{ \eta_{i}\}=0})^{2}\bigg] \nonumber \\&=&- \frac{1}{4}[\sum_{n,m}(W_{n,n+1}W_{m,m+1}-W_{n,m}W_{n+1,m+1}+W_{n,m}W_{n+1,m+1}) - (\sum_{n}W_{n,n+1})^{2}] \nonumber\\
&=& - \frac{1}{4}[N\sum_{n}(W_{1}^2-W_{n}^{2}+W_{n+1}W_{n-1}) - N^{2}W_{1}^{2}] \nonumber\\
&=&\frac{1}{4}N\sum_{n}(W_{n}^{2}-W_{n+1}W_{n-1}).
\eeqa
Note that the expression on the right hand side of the final equality of (\ref{xxx}) is also valid for the zero-order term, i.e. when $n=m$ and so $W_{0}=1$. Taking this out of the summation and using the symmetry of the ring, (taking N even for convenience) we have
\begin{equation}
\begin{aligned}
\kappa_{2}&=\frac{1}{4}N\sum_{n}(W_{n}^{2}-W_{n+1}W_{n-1})\\
&=\frac{1}{4}N(1 - W_{1}^{2} + 2\sum_{n = 1}^{N/2}(W_{n}^{2}-W_{n+1}W_{n-1})).
\end{aligned}
\end{equation}
In the case of $W_{n}$ being described by an expression of the form (\ref{corrn}), the resulting limiting expression for $\kappa_{2}$ is then
\begin{equation}
\kappa_{2} = \frac{N}{2}C\tau_{Q}^{-\nu}. 
\end{equation}

\subsection{Explicit Calculation of $\kappa_{3}$}
Returning to the expression for the third cumulant
\beqa
\kappa_{3} &=& -\frac{1}{8}\Bigg[\sum_{n,m,l}(\partial^{6}_{l+1,l,m+1,m,n+1,n}\Psi)|_{\{ \eta_{i}\}=0} - 3\left(\sum_{n}(\partial^{2}_{n+1,n}\Psi)|_{\{ \eta_{i}\}=0}\right)
\left(\sum_{n}(\partial^{2}_{m+1,m,n+1,n}\Psi)|_{\{ \eta_{i}\}=0}\right)\nonumber \\
 & &+2(\sum_{n}(\partial^{2}_{n+1,n}\Psi)|_{\{ \eta_{i}\}=0})^{3}\Bigg].
\eeqa
Making use of  translational invariance in the system allows us to dispense with the first indices, defining the $n$'th and $m$'th spins relative to their distance from the first:
\begin{align}
\kappa_{3} = -\frac{1}{8}\left[N\sum_{n,m=0}^{N-1}(\partial^{6}_{m+1,m,n+1,n,1,0}\Psi)|_{\{ \eta_{i}\}=0} - 3N^{2}W_{1}(\sum_{n=0}^{N-1}(\partial^{2}_{n+1,n,1,0}\Psi)|_{\{ \eta_{i}\}=0})+2N^{3}W_{1}^{3}
\right] 
\end{align}
A lengthy differentiation process of the terms containing $\Psi$, or equivalently collecting all index pairs in a Fermionic Wick contraction yields:
\beqa
\kappa_{3} &=& -\frac{1}{8}[N\sum_{n,m=0}^{N-1}(W_{1}^{3}+W_{1}(W_{n+1}W_{n-1}-W_{n}^{2}+W_{m+1}W_{m-1}-W_{m}^{2}+W_{n-m+1}W_{n-m-1}-W_{n-m}^{2})\nonumber\\ 
& &+ W_{n-m}(W_{n}W_{m-1}-W_{m}W_{n-1}+W_{n+1}W_{m}-W_{m+1}W_{n}) + W_{n-m+1}(W_{m+1}W_{n-1}-W_{n}W_{m})
\nonumber \\
 & &+W_{n-m-1}(W_{n}W_{m}-W_{n+1}W_{m-1}))-3N^{2}W_{1}\sum_{n=0}^{N-1}(W_{1}^{2}-W_{n}^{2}+W_{n+1}W_{n-1})+2N^{3}W_{1}^{3}
]. 
\eeqa
Terms proportional to $W_{1}^{3}$ immediately cancel, as do those symmetric under the transformations $n\rightarrow -n$ and $n\leftrightarrow m$, with an analogous counterpart with opposite sign (i.e. the second, third and fourth brackets within the first summation), leading to
\beqa
\kappa_{3} &=& -\frac{1}{8}NW_{1}\Bigg[\sum_{n,m=0}^{N-1}(W_{n+1}W_{n-1}-W_{n}^{2}+W_{m+1}W_{m-1}-W_{m}^{2}+W_{n-m+1}W_{n-m-1}-W_{n-m}^{2})\nonumber\\ & & -3N\sum_{n=0}^{N-1}(W_{n+1}W_{n-1}-W_{n}^{2})
\Bigg]. 
\eeqa
Collecting alike terms in the second summation yields
\begin{equation}
\kappa_{3} = -\frac{1}{8}NW_{1}\left[\sum_{n,m=0}^{N-1}(W_{n-m+1}W_{n-m-1}-W_{n-m}^{2})-N\sum_{n=0}^{N-1}(W_{n+1}W_{n-1}-W_{n}^{2})\right]. 
\end{equation}
Making use of the periodic boundary conditions to reduce the sum, we find
\begin{equation}
\kappa_{3} = -\frac{1}{8}NW_{1}\left[4\sum_{n,m=0}^{(N-1)/2}(W_{n-m+1}W_{n-m-1}-W_{n-m}^{2})-2N\sum_{n=0}^{(N-1)/2}(W_{n+1}W_{n-1}-W_{n}^{2})\right]. 
\end{equation}
For a further simplification, it proves convenient to redefine indices as $l=|n-m|$, to find that there are $(N+1)/2$ terms in which $l=0$, $2((N+1)/2-1)=(N-1)$ terms such that $l=1$, $2((N+1)/2-2)=(N-3)$ with $l=2$, etc. With this redefinition, we can write
\beqa
\kappa_{3} 
= -\frac{1}{8}NW_{1}\left[2(N+1)(W_{1}^{2}-1)
+4\sum_{l=1}^{(N-1)/2}(N+1-2l)(W_{l+1}W_{l-1}-W_{l}^{2})
-2N\sum_{n=0}^{(N-1)/2}(W_{n+1}W_{n-1}-W_{n}^{2}) \right].  \nonumber\\
\eeqa
Evaluating the first term in the second summation, and collecting alike terms gives
\beqa
\kappa_{3} = -\frac{1}{4}NW_{1}\left[W_{1}^{2}-1+\sum_{l=1}^{(N-1)/2}(N+2-4l)(W_{l+1}W_{l-1}-W_{l}^{2}))\right],
\eeqa
where we recognize the first two terms as $\kappa_{3}=p(1-p)(1-2p)=\frac{1}{4}W_{1}(1-W_{1}^{2})$ for $p=\frac{1}{2}(1-W_{1})$. Once again, if in the $\tau_{Q}\rightarrow \infty$ limit  Eq. (\ref{corrn}) holds, we have
\begin{equation}
\kappa_{3}= \frac{N}{2}C\tau_{Q}^{-\nu}.
\end{equation}

\section{Calculation of the 2-Point Correlator}\label{2PCorrApp}
In this appendix, we detail the computation of the two-point correlator $W_{n}(\tau_{Q})$ in Eq. (\ref{intWn}) for the Ising model under Glauber dynamics and slow quenches. Before dwelling on specific cases, we note that by applying the recursion formula for modified Bessel functions 
\begin{equation}
I_{n-1}(\eta)-I_{n+1}(\eta) =\frac{2n}{\eta}I_{n}(\eta), 
\end{equation}
 the integral $W_{n}(\tau_{Q})$ to be written as
\begin{equation}
W_{n}(\tau_{Q})=n\int_{0}^{2\tau_{Q}(1-\frac{A}{1+\alpha})}\mathrm{d}\eta e^{-\eta-\frac{A}{1+\alpha}\frac{\eta^{1+\alpha}}{(2\tau_{Q})^{\alpha}}}\frac{I_{n}(\eta)}{\eta}, \quad n \geq 1.
\end{equation}

\subsection{Linear Cooling}
In the case of $\alpha=1$, the integral reduces to
\begin{equation}
W_{n}(\tau_{Q})=n\int_{0}^{\tau_{Q}}\mathrm{d}\eta e^{-\eta-\frac{\eta^{2}}{4\tau_{Q}}}\frac{I_{n}(\eta)}{\eta}.
\end{equation}
Introducing $x=\eta/(2\sqrt{\tau_{Q}})$,  the integral becomes
\begin{equation}
W_{n}(\tau_{Q})=n\int_{0}^{\sqrt{\tau_{Q}}/2}\mathrm{d}x e^{-2\sqrt{\tau_{Q}}x-x^{2}}\frac{I_{n}(2\sqrt{\tau_{Q}}x)}{x}.
\end{equation}
Split the integral into two parts, and define $f=f(\tau_{Q})$ a function to be optimized later
\begin{equation}
n\left[\int_{0}^{f(\tau_{Q})}\mathrm{d}x  e^{-2\sqrt{\tau_{Q}}x-x^{2}}\frac{I_{n}(2\sqrt{\tau_{Q}}x)}{x} +\int_{f(\tau_{Q})}^{\sqrt{\tau_{Q}}/2}\mathrm{d}x e^{-2\sqrt{\tau_{Q}}x-x^{2}}\frac{I_{n}(2\sqrt{\tau_{Q}}x)}{x}\right].
\end{equation}
The Taylor and asymptotic expansions of the modified Bessel function $I_{n}(x)$ are given by
\begin{equation}
\label{expansions}
\begin{aligned}
I_{\nu}(z)&= \sum_{k=0}^{\infty}\frac{1}{\Gamma(k+\nu+1)k!}\left(\frac{z}{2}\right)^{2k+\nu}, \\
I_{\nu}(z) &\sim \frac{e^{z}}{\sqrt{2\pi z}}\sum_{k=0}^{\infty}(-1)^{k}\frac{a_{k}(\nu)}{z^{k}} \ \mathrm{as} \ z\rightarrow \infty.
\end{aligned}
\end{equation}
where $a_{k}(\nu)$ denotes a member of the class of polynomials defined by the general formula
\begin{equation}
a_{n}(\nu) = \frac{(4\nu^{2}-1)(4\nu^{2}-9)...(4\nu^{2}-(2n-1)^{2})}{8^{n}\Gamma(n+1)}.
\end{equation}
Plugging the upper expression (\ref{expansions}) into the lower integral, the first term becomes
\beqa
 n\sum_{k=0}^{\infty}\frac{\tau_{Q}^{k+\frac{n}{2}}}{\Gamma(k+n+1)k!}\int_{0}^{f(\tau_{Q})}\mathrm{d}x e^{-2\sqrt{\tau_{Q}}x-x^{2}}x^{2k+n-1}  \approx n\sum_{k=0}^{\infty}\frac{\tau_{Q}^{k+\frac{n}{2}}}{k!}\int_{0}^{f(\tau_{Q})}\mathrm{d}x e^{-2\sqrt{\tau_{Q}}x}x^{2k+n-1},
\eeqa
where the approximation is justified pre-emptively by the choice of $f(\tau)$ , namely that it should go to zero in the limit of large $\tau$. In such a limit, we have that $\lim_{x\rightarrow 0}\frac{e^{-\sqrt{2\tau_{Q}}x-x^{2}}}{e^{-\sqrt{2\tau_{Q}}x}}=\lim_{x\rightarrow 0}e^{-x^{2}}=1$. Solving the integral exactly, we find it to be equal to
\begin{equation}
\label{lowend}
\begin{aligned}
&n\sum_{k=0}^{\infty}\frac{\tau_{Q}^{k+\frac{n}{2}}}{\Gamma(k+n+1)k!}\frac{1}{2^{2k+n}\tau_{Q}^{k+\frac{n}{2}}} \left[ \Gamma(2k+n)-\Gamma(2k+n,2f(\tau_{Q})\sqrt{\tau_{Q}}) \right]\\
&=n\sum_{k=0}^{\infty}\frac{1}{\Gamma(k+n+1)k!}\frac{1}{2^{2k+n}} \left[ \Gamma(2k+n)-\Gamma(2k+n,2f(\tau_{Q})\sqrt{\tau_{Q}}) \right],
\end{aligned}
\end{equation}
where $\Gamma(a,b)$ denotes the upper incomplete gamma function. Turning our attention to the second integral, and making use of the asymptotic form (\ref{expansions}) we find it to be  given by
\begin{equation}
\label{asymp}
\frac{n}{2\sqrt{\pi}\tau^{\frac{1}{4}}}\sum_{k=0}^{\infty}\frac{(-1)^{k}}{(2\sqrt{\tau_{Q}})^{k}}a_{k}(n)\int_{f(\tau_{Q})}^{\sqrt{\tau_{Q}}/2}\mathrm{d}x\frac{e^{-x^{2}}}{x^{k+\frac{3}{2}}}.
\end{equation} 
The integral part of the expression (\ref{asymp}) is exactly solvable, and the resulting form is
\begin{equation}
-\frac{n}{4\sqrt{\pi}\tau^{\frac{1}{4}}}\sum_{k=0}^{\infty}\frac{(-1)^{k}}{(2\sqrt{\tau_{Q}})^{k}}a_{k}(n)\left[ \Gamma(-\frac{k}{2}-\frac{1}{4},\frac{\tau_{Q}}{4})- \Gamma(-\frac{k}{2}-\frac{1}{4},f(\tau_{Q})^{2})\right].
\end{equation} 
Examining the first sum in (\ref{lowend}), we find:
\begin{equation}
n\sum_{k=0}^{\infty}\frac{\Gamma(2k+n)}{\Gamma(k+n+1)k!}\frac{1}{2^{2k+n}} = n\frac{\Gamma(n)}{\Gamma(n+1)} = 1,
\end{equation}
leading to the full expression for $W_{n}$
\beqa
W_{n}&=&1 - n\sum_{k=0}^{\infty}\frac{1}{\Gamma(k+n+1)k!}\frac{1}{2^{2k+n}}\Gamma(2k+n,2f(\tau_{Q})\sqrt{\tau_{Q}})\nonumber \\& &-\frac{n}{4\sqrt{\pi}\tau^{\frac{1}{4}}}\sum_{k=0}^{\infty}\frac{(-1)^{k}}{(2\sqrt{\tau_{Q}})^{k}}a_{k}(n)\left[ \Gamma(-\frac{k}{2}-\frac{1}{4},\frac{\tau_{Q}}{4}) - \Gamma(-\frac{k}{2}-\frac{1}{4},f(\tau_{Q})^{2})\right]. 
\eeqa
Now we go about optimizing $f(\tau_{Q})$. We know that $\lim_{\tau_{Q} \rightarrow \infty}W_{n}(\tau_{Q})=1$, since an infinite quench has thermal motion allowing spins to align, and continued coarsening dynamics to ensure that any excitation in the system is eventually removed. This requires that $f(\tau_{Q})\sqrt{\tau_{Q}}\rightarrow \infty$ as $\tau_{Q}\rightarrow\infty$. Furthermore, we wish to have an expression with complete $\Gamma$ functions, which do not depend on $\tau_{Q}$. Thus, in the limit of $\tau_{Q} \rightarrow \infty$, this leads to the condition that $f(\tau_{Q})\rightarrow 0$ as $\tau_{Q} \rightarrow \infty$. Thus, we pick as a suitable choice $f(\tau_{Q})=\tau_{Q}^{-\frac{1}{4}}$. We see upon expansion in inverse powers of $\tau_{Q}$ that the second and third $\Gamma$ functions are exponentially suppressed in powers of $\tau_{Q}$. Applying these conditions while taking the leading power in $\tau_{Q}$, and thereafter simplifying the $\Gamma$ function using standard identities leaves us with the final expression:
\begin{equation}
W_{n} =1-\frac{n}{\sqrt{\pi}\tau_{Q}^{\frac{1}{4}}}\left[ \Gamma\left(\frac{3}{4}\right)-\frac{1}{6\sqrt{\tau_{Q}}}a_{1}(n)\Gamma\left(\frac{1}{4}\right)+\mathcal{O}(\tau_{Q}^{-1})\right].
\end{equation}
In the limit of slow cooling, T final density of defects is governed by the  power-law behavior
\begin{equation}
\rho(\tau_{Q}) =\frac{\kappa_1}{N}
=
\frac{\Gamma(\frac{3}{4})}{2\sqrt{\pi}\tau_{Q}^{\frac{1}{4}}}. 
\end{equation}

\subsection{Algebraic Cooling}
We turn our attention now to the general case of algebraic cooling. Restating the integral (\ref{algWn})
\begin{equation}
W_{n}(\tau_{Q})=n\int_{0}^{2\tau_{Q}(1-\frac{A}{1+\alpha})}\mathrm{d}\eta e^{-\eta-\frac{A}{1+\alpha}\frac{\eta^{1+\alpha}}{(2\tau_{Q})^{\alpha}}}\frac{I_{n}(\eta)}{\eta}.
\end{equation}
We proceed by making the substitution $x=c\eta(2\tau_{Q})^{-\frac{\alpha}{1+\alpha}}$, where $c=(\frac{A}{1+\alpha})^{\frac{1}{1+\alpha}}$ is defined for convenience. Applying this substitution, and defining $c'=(\frac{A}{1+\alpha})^{\frac{1}{1+\alpha}}(1-\frac{A}{1+\alpha})$, we find:
\begin{equation}
W_{n}=n\int_{0}^{c'(2\tau_{Q})^{1-\frac{\alpha}{1+\alpha}}}\mathrm{d}xe^{-x(2\tau_{Q})^{\frac{\alpha}{1+\alpha}}/c+x^{2}}\frac{I_{n}(\frac{x(2\tau_{Q})^{\frac{\alpha}{1+\alpha}}}{c})}{x}.
\end{equation}
Splitting the integral in the same fashion as before, we find that the lower contribution becomes approximately equal to
\beqa
n\sum_{k=0}^{\infty}\frac{1}{\Gamma(k+n+1)k!}\frac{(2\tau_{Q})^{\frac{\alpha}{1+\alpha}(2k+n)}}{(2c)^{2k+n}} \int_{0}^{f(\tau_{Q})}\mathrm{d}xe^{-\frac{x(2\tau_{Q})^{\frac{\alpha}{1+\alpha}}}{c}}x^{2k+n-1}
\eeqa
while the upper contribution is
\beqa
\frac{n}{\sqrt{2\pi/c}(2\tau_{Q})^{\frac{\alpha}{2(1+\alpha)}}}\sum_{k=0}^{\infty}\frac{(-1)^{k}}{c^{-k}(2\tau_{Q})^{\frac{k\alpha}{1+\alpha}}}a_{k}(n)
\int_{f(\tau_{Q})}^{c'(\tau_{Q})^{1-\frac{\alpha}{1+\alpha}}}\mathrm{d}x\frac{e^{-x^{2}}}{x^{k+\frac{3}{2}}}.
\eeqa
As in the previous section, we find conditions on $f(\tau)$, which turn out to be $\lim_{\tau\rightarrow \infty}\tau_{Q}^{\frac{\alpha}{1+\alpha}}f(\tau_{Q})=\infty$ and $\lim_{\tau \rightarrow \infty}f(\tau_{Q})=0$. Thus defining $f(\tau_{Q}) = \tau_{Q}^{-\frac{\alpha}{2(1+\alpha)}}$, and taking the limit as before, we find that:
\begin{equation}
W_{n} = 1 - n\sqrt{\frac{2c(\alpha)}{\pi}}\frac{1}{(2\tau_{Q})^{\frac{\alpha}{2(1+\alpha)}}}\left[\Gamma\left(\frac{3}{4}\right) -\frac{c(\alpha)a_{1}(n)}{3(2\tau_{Q})^{\frac{\alpha}{1+\alpha}}}\Gamma\left(\frac{1}{4}\right)+\mathcal{O}\left(\tau_{Q}^{-\frac{2\alpha}{1+\alpha}}\right)\right],
\end{equation}
giving the final density of defects equal to
\begin{equation}
\rho(\tau_{Q})=\frac{\kappa_{1}}{N}= \sqrt{\frac{c(\alpha)}{2\pi}}\frac{\Gamma(\frac{3}{4})}{(2\tau_{Q})^{\frac{\alpha}{2(1+\alpha)}}},
\end{equation}
to leading order.

\subsection{Exponential Cooling}
In the case of exponential cooling, we begin with equation (37)
\begin{equation}
W_{n}=n\int_{0}^{\tau_{Q}}\mathrm{d}t(1-B\exp(-\frac{b}{(1-t/\tau_{Q})^{\beta}}))\exp(2(\tau_{Q}-t))\frac{I_{n}(h(\tau_{Q},t))}{h(\tau_{Q},t)},
\end{equation}
where, in this case,
\begin{equation}
h(\tau_{Q},t) = \tau_{Q}-t - B \int_{t}^{\tau_{Q}}\mathrm{d}t'\exp(-\frac{b}{(1-t/\tau_{Q})^{\beta}}).
\end{equation}
Using the substitution $u=\frac{b^{1/\beta}}{(1-t/\tau_{Q})}$, the integral becomes.
\begin{equation}
\int_{t}^{\tau_{Q}}\mathrm{d}t'\exp(-\frac{b}{(1-t/\tau_{Q})^{\beta}}) =b^{1/\beta}\tau_{Q}\int_{b^{1/\beta}/(1-t/\tau_{Q})}^{\infty}\frac{\exp(-u^{\beta})}{u^{2}}.
\end{equation}
The integral on the left hand side of (75) admits an analytical solution in terms of the incomplete gamma function as
\begin{equation}
b^{1/\beta}\tau_{Q}\int_{b^{1/\beta}/(1-t/\tau_{Q})}^{\infty}\frac{\exp(-u^{\beta})}{u^{2}} = b^{1/\beta}\tau_{Q}\left[-\frac{\Gamma(-\frac{1}{\beta},u^{\beta})}{\beta} \right]_{u=b^{1/\beta}/(1-t/\tau_{Q})}^{u=\infty},
\end{equation}
giving
\begin{equation}
b^{1/\beta}\tau_{Q}\int_{b^{1/\beta}/(1-t/\tau_{Q})}^{\infty}\frac{\exp(-u^{\beta})}{u^{2}} = \frac{b^{1/\beta}\tau_{Q}}{\beta}\Gamma(-\frac{1}{\beta},b/(1-t/\tau_{Q})^{\beta}).
\end{equation}
Therefore, we have that
\begin{equation}
h(\tau_{Q},t) = \tau_{Q}-t - B\frac{b^{1/\beta}\tau_{Q}}{\beta}\Gamma(-\frac{1}{\beta},b/(1-t/\tau_{Q})^{\beta}).
\end{equation}
Defining then $\eta = 2h(\tau_{Q},t)$, we have:
\begin{equation}
\frac{\mathrm{d}\eta}{\mathrm{d}t}=-2(1-B\exp(-b/(1-t/\tau_{Q}))) = -2\gamma(t).
\end{equation}
Inverting as in \cite{Krapivsky10}, we find that:
\begin{equation}
1-\frac{t}{\tau_{Q}}=\frac{\eta}{2\tau_{Q}}+\frac{B}{b\beta}\left(\frac{\eta}{2\tau_{Q}}\right)^{1+\beta}\exp\left\{-b\left( \frac{\eta}{2\tau_{Q}}\right)^{-\beta}\right\},
\end{equation}
and so the integral becomes
\begin{equation}
\label{etaintegral2}
n\int_{0}^{2\tau_{Q}(1-\frac{Bb^{1/\beta}}{\beta}\Gamma(-1/\beta,b))}\mathrm{d}\eta\exp\left\{-\eta- \frac{B}{b\beta}\left(\frac{\eta}{2\tau_{Q}}\right)^{\beta}\eta\exp\left\{ -b\left(\frac{\eta}{2\tau_{Q}}\right)^{-\beta}\right\} \right\}\frac{I_{n}(\eta)}{\eta}.
\end{equation}
Splitting the integral (\ref{etaintegral2}) into an upper and lower part, while defining $\eta=2\tau_{Q}(\xi b/\ln(\tau_{Q}))^{1/\beta}$ to substitute in the upper contribution, while keeping the first contribution in terms of $\eta$ for convenience gives
\begin{eqnarray}
& & \label{lowerexp}
n\sum_{k=0}^{\infty}\frac{1}{\Gamma
(k+n+1)k!}\frac{1}{2^{2k+n}}\int_{0}^{\eta\left(f(\tau_{Q})\right)}\mathrm{d}\eta e^{-\eta}\eta^{2k+n-1} +\frac{n}{\beta} \frac{1}{2\sqrt{\pi\tau_{Q}}}\left(\frac{\ln(\tau_{Q})}{b}\right)^{1/2\beta}\nonumber \\ & &\times\int_{f(\tau_{Q})}^{\frac{\ln(\tau_{Q})}{b}c}\frac{\mathrm{d}\xi}{\xi^{1+\frac{1}{2\beta}}}  \exp\left\{ -\frac{2B \xi}{\beta}\left( \frac{\xi b}{\ln(\tau_{Q})}\right)^{1/\beta} \tau_{Q}^{1-\frac{1}{\xi}} \right\}   ,
\end{eqnarray}
where we have defined $c=(1-\frac{Bb^{1/\beta}}{\beta}\Gamma(-1/\beta,b))^{\beta}$ for convenience. Taking a cue from \cite{Krapivsky10} and noting that the upper part of the integral converges to zero for all $\xi > 1$, we find that its contribution can be replaced by
\begin{equation}
\label{zetaint}
\frac{n}{\beta} \frac{1}{2\sqrt{\pi\tau_{Q}}}\left(\frac{\ln(\tau_{Q})}{b}\right)^{1/2\beta}\int_{f(\tau_{Q})}^{1}\frac{\mathrm{d}\xi}{\xi^{1+\frac{1}{2\beta}}}.
\end{equation}
Solving the integral (\ref{zetaint}) leads to the expression
\begin{equation}
\label{zetaintsolved}
n \frac{1}{\sqrt{\pi\tau_{Q}}}\left(\frac{\ln(\tau_{Q})}{b}\right)^{\frac{1}{2\beta}}\left[ \frac{1}{f(\tau_{Q})^{\frac{1}{2\beta}}}- 1 \right].
\end{equation}
The condition implied by both (\ref{zetaintsolved}) to remove the divergence and the expansion of the lower expression (\ref{lowerexp}) is that
\begin{equation}
\lim_{\tau_{Q}\rightarrow \infty} \frac{f(\tau_{Q})\tau_{Q}^{\beta}}{\ln(\tau_{Q})} = \infty,
\end{equation}
while still $\lim_{\tau_{Q}\rightarrow\infty}f(\tau_{Q})=0$. Thus, we may pick $f(\tau_{Q})=\ln(\tau_{Q})\tau_{Q}^{1-\beta}$, and find the final expression for exponential cooling to be
\begin{equation}
W_{n}=1- n \frac{1}{\sqrt{\pi\tau_{Q}}}\left(\frac{\ln(\tau_{Q})}{b}\right)^{\frac{1}{2\beta}}.
\end{equation}
The density of defects is, therefore
\begin{equation}
\rho(\tau_{Q})=\frac{\kappa_{1}}{N} = \frac{1}{2\sqrt{\pi\tau_{Q}}}\left(\frac{\ln(\tau_{Q})}{b}\right)^{\frac{1}{2\beta}}.
\end{equation}

\begin{figure}[t]
	\includegraphics[width=0.7\linewidth]{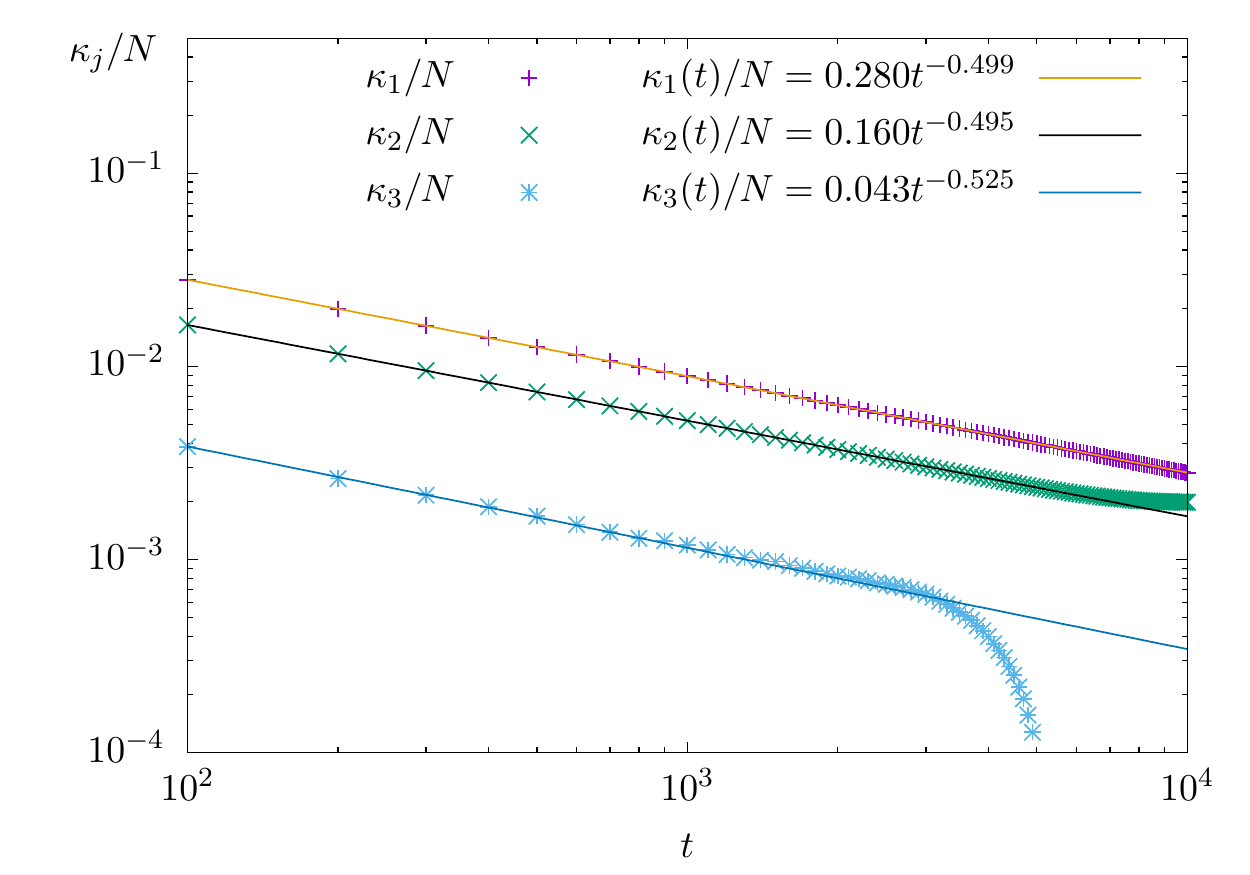}
	\caption{First three cumulants versus time after a sudden quench to $T = 0$ of 1D Ising chain obtained from Glauber dynamics simulations. The system size is $N= 500$. For all three cumulants, each data point is obtained by averaging over $M = 500000$ independent simulations.}
		\label{fig:cumulants_coarsening}
\end{figure}

\section{Coarsening dynamics of 1D Ising chain}

\label{sec:coarsening}

In this section, we present Glauber dynamics simulations of the coarsening phenomenon in 1D Ising model.  Coarsening, or phase-ordering dynamics, underlies numerous natural processes including phase separation, grain growth, and biological pattern formation~\cite{bray02coarsening}. It is generally believed that the ordering process following the quench of a system from an initial state at high temperature to a final state below the critical point obeys dynamic scaling in the asymptotic time regime~\cite{bray02coarsening}. The intuitive argument for this dynamical scaling is that at the late stage of phase ordering, the typical domain size $L(t)$ is the only important length scale in the system, and any time dependence takes place through $L(t)$. The standard picture is that the growth of typical domain size follows a power-law $L(t) \sim t^{1/z}$, where the dynamical exponent $z$ is usually independent of details of the system and even the spatial dimensions. On the other hand, similar to critical phenomena, the dynamical scaling of coarsening can be classified into universality classes that depend on the symmetry of the order parameters, whether the order parameter is conserved or not, and coupling to other dynamical variables. For example, in dimensions greater than or equal to 2, coarsening of Ising-like domains is described by an exponent $z = 2$ for non-conserved order parameters, and $z = 3$ for conserved ones. 

In one dimension, since domains of ordered spins are sandwiched by two kinks, the typical domain size is related to the density of kinks, or first cumulant, via $L(t) \sim N/\kappa_1(t)$. Consequently, if dynamical scaling also holds for quench to the $T = 0$ critical point, we expect a power-law behavior for the first cumulant, $\kappa_1(t) \sim t^{-1/z}$. Fig.~\ref{fig:cumulants_coarsening} shows the time dependence of the first three cumulants after a sudden quench to $T = 0$ obtained from our Glauber dynamics simulations. It can be seen that not only $\kappa_1$, but all three cumulants can be well described by a scaling relation $\kappa_j \sim t^{-1/2}$. This result seems to be consistent with the prediction that $z = 2$ for coarsening of a non-conserved Ising-type order parameter, which is indeed the case for the Glauber dynamics. However, the general $L \sim t^{1/2}$ scaling, also known as the Allen-Cahn law in high dimensions, originates from a domain growth in which the linear growth rate is proportional to the curvature of the interface. For 1D Ising chain the $z=2$ exponent, on the other hand, comes from the random walks of kinks. It is worth noting that at $T = 0$, a kink can move to either the left or right lattice points with equal probability in the Glauber simulation. Since the root mean square displacement of such random walker is $\langle \Delta \ell \rangle \sim \sqrt{t}$, two kinks within this distance will be annihilated, thus increasing the size of ordered domains.

\end{widetext}
\end{appendix}

\newpage

\bibliography{KinkFCSClassBib}

\end{document}